# A Review of Applications, Prospects, and Challenges of Proton-Conducting Zirconates in Electrochemical Hydrogen Devices

**M. Khalid Hossain [1,2,\*], S. M. Kamrul Hasan [3], M. Imran Hossain [4], Ranjit. C. Das [5], H. Bencherif [6], M. H. K. Rubel [7], Md. Ferdous Rahman [8], Tanvir Emrose [9] and Kenichi Hashizume [1,\*]**

[1]  Department of Advanced Energy Engineering Science, IGSES, Kyushu University, Fukuoka 816-8580, Japan
[2]  Institute of Electronics, AERE, Bangladesh Atomic Energy Commission, Dhaka 1349, Bangladesh
[3]  Department of Mechanical Engineering, Auburn University, Auburn, AL 36849, USA
[4]  Institute for Micromanufacturing, Louisiana Tech University, Ruston, LA 71270, USA
[5]  Materials Science and Engineering, Florida State University, Tallahassee, FL 32306, USA
[6]  Higher National School of Renewable Energies, Environment and Sustainable Development,
    Batna 05078, Algeria
[7]  Department of Materials Science and Engineering, University of Rajshahi, Rajshahi 6205, Bangladesh
[8]  Department of Electrical and Electronic Engineering, Begum Rokeya University, Rangpur 5400, Bangladesh
[9]  School of Electrical Engineering and Computer Science, Louisiana State University,
    Baton Rouge, LA 70803, USA
\*  Correspondence: khalid.baec@gmail.com and khalid@kyudai.jp (M.K.H.) ORCID: https://orcid.org/0000-0003-4595-6367

**Abstract:** In the future, when fossil fuels are exhausted, alternative energy sources will be essential for everyday needs. Hydrogen-based energy can play a vital role in this aspect. This energy is green, clean, and renewable. Electrochemical hydrogen devices have been used extensively in nuclear power plants to manage hydrogen-based renewable fuel. Doped zirconate materials are commonly used as an electrolyte in these electrochemical devices. These materials have excellent physical stability and high proton transport numbers, which make them suitable for multiple applications. Doping enhances the physical and electronic properties of zirconate materials and makes them ideal for practical applications. This review highlights the applications of zirconate-based proton-conducting materials in electrochemical cells, particularly in tritium monitors, tritium recovery, hydrogen sensors, and hydrogen pump systems. The central section of this review summarizes recent investigations and provides a comprehensive insight into the various doping schemes, experimental setup, instrumentation, optimum operating conditions, morphology, composition, and performance of zirconate electrolyte materials. In addition, different challenges that are hindering zirconate materials from achieving their full potential in electrochemical hydrogen devices are discussed. Finally, this paper lays out a few pathways for aspirants who wish to undertake research in this field.

**Keywords:** perovskite oxide; proton-conducting oxide; zirconate; electrochemical device; tritium monitoring; tritium recovery; hydrogen sensors; hydrogen pumps

## 1. Introduction

As a result of the Industrial Revolution and technological advancements, the globe requires alternative energy sources to supply the ever-increasing demand for energy [1–3]. In addition, With the rapid depletion of fossil fuel resources and the negative impact of fossil fuel combustion on our environments [4–7], scientists have turned their attention to other renewable sources, such as electrochemical hydrogen devices based on proton-conducting materials [8–11]. Proton conductors typically have positively charged protonic species, such as $H^+$, $H_3O^+$, and $NH_4^+$ [12,13]. Proton-conducting materials provide higher conductivity at lower temperatures with longer lifetimes and less expense than traditional oxide ionic electrolyte conductors [14,15]. In addition, these conductors lose conductivity at higher temperatures due to reversible or irreversible loss of carriers [16]. These characteristics enable these materials to operate at narrow ranges of temperature.

Proton conductors can be used in various electrochemical energy devices, such as batteries, fuel-cell electrolytes, water electrolyzers' membrane, hydrogen pumps, hydrogen sensors, and hydrogen gas separation systems [17–20]. Organic polymer, inorganic oxides, and lattice defect oxides are examples of the different types of proton conductors. Compared to the other proton conductors, lattice defect-type oxides, i.e., perovskite-type proton-conducting oxides, are the promising proton conductors due to having the highest proton conductivity and chemical stability within desired temperatures [12,21,22]. A typical chemical formula of a perovskite proton conductor is $ABO_3$ (A = Ba, Ca, Sr, etc.; B = Zr, Ce, Tb, Th, etc.) [23,24]. In addition, perovskite materials have higher conversion efficiency and are less expensive than other proton conductors [25,26]. These unique properties of perovskite materials have increased their utility in renewable energy applications, especially in solar cells [27]. Among different types of perovskite proton-conducting materials, zirconate materials are the most widely studied/used due to their high chemical stability and excellent proton conductivity [16,28–30].

Zirconate materials such as $BaZrO_3$-based materials are considered promising proton-conducting materials and are widely used in chemical and electrical sectors. However, many studies have shown that cerate-based proton conductors such as $BaCeO_3$ have high proton conductivity among perovskite-based materials [12]. The drawback of $BaCeO_3$-based materials is that they are unstable in $CO_2$ and water vapor atmospheres, making them unsuitable for applications [31,32]. In contrast, $BaZrO_3$-based proton conductors are stable in $CO_2$ and water vapor environments which are attractive properties for electrochemical device application in harsh atmospheres [25]. Moreover, $BaZrO_3$-based materials have better physical properties, including chemical stability and higher mechanical hardness than $BaCeO_3$-based proton-conducting material [33]. Ken Kurosaki et al. reported that $BaZrO_3$ exhibits high thermal conductivity due to the high strength between Zr and O [34]. However, the $BaZrO_3$-based proton conductor's proton conductivity is lower than the $BaCeO_3$-based proton conductor, which can be improved by doping with trivalent cations such as $Gd^{3+}$, $Y^{3+}$, $In^{3+}$, $Yb^{3+}$ [35,36]. Pergolesi et al. have reported that $Y^{3+}$ doped in $BaZrO_3$ enhances chemical stability, but the poor sinterability increases grain-boundary resistance, which is responsible for reducing proton conductivity [37]. Therefore, the sintering temperature must be increased with decreased grain-boundary resistance to improve electrical properties in zirconate-based proton conductors [38]. Recent research has shown that In-doped zirconate-based perovskite proton conductors exhibit better sintering activities with excellent chemical stability [39]. Consequently, experiments with different doping concentrations and synthesis methods are used to develop high-performing doped $BaZrO_3$ material.

Zirconate materials have low thermal conductivity, low dielectric loss, and very low thermal expansion coefficient [16,40,41], making them more favorable for electrochemical devices than other proton-conducting oxide materials. Furthermore, compared to other proton-conducting materials in hydrogen sensors, zirconate-based hydrogen sensors have been demonstrated to be affordable, portable, and temporally correct due to their high chemical stability, smaller dimensions, and cheapness [16,42]. Hydrogen can be separated in zirconate-based proton conductors in a controlled way simply by changing the applied current in the electrochemical cell; thus, they can be utilized as hydrogen pumps [16]. Zirconate proton conductors can be used as membrane separators at high temperatures, enabling them to act as a sensitive tritium monitor system [43]. Such a device is helpful in removing inference from radionucleotides and concentrating tritium, since it can operate like an electrochemical hydrogen isotope pump [43]In addition, tritium release has been reported in zirconate proton-conducting material spheres as far back as 30 years ago, and scientists are making more advancements in that technology [44–49]. Research has established the reputation of zirconate materials as an efficient tritium recovery system. As a result, zirconate materials have garnered significant interest in the scientific community, and they have published numerous articles about their work. However, none of the papers go into depth about the utility of zirconium proton conductors in electrochemical device applications.

This comprehensive review discusses various aspects and potential of zirconate proton conductors in electrochemical device applications. The first section presents a short overview of proton-conducting zirconates and electrochemical hydrogen devices, including tritium monitoring systems, tritium recovery systems, hydrogen sensors, and hydrogen pumps. In the following section, we broadly discuss recent developments and potential of zirconate-based proton conducting materials used as an electrolyte in the application of electrochemical devices. The prospects and challenges of zirconate proton conductors in electrochemical applications are discussed in the final section of the review (Figure 1) based on the published literature and the authors' experience.

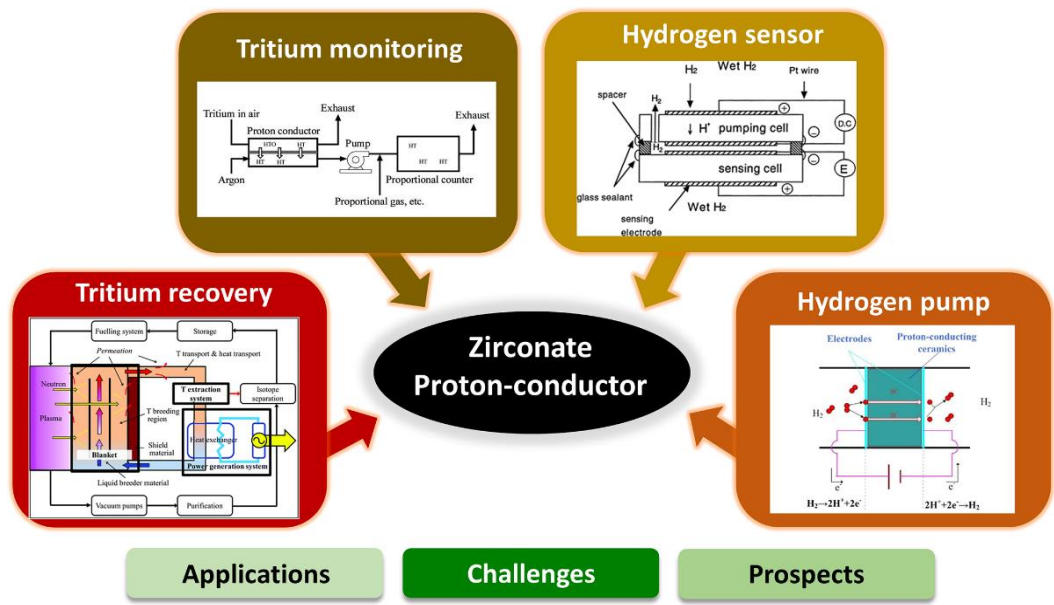

**Figure 1.** An outline of the main points of this review shown schematically.

## 2. Proton-Conducting Zirconates

Perovskite proton-conductor oxides, i.e., zirconates and cerate-based materials, are well-known proton conductors for electrochemical device applications due to their excellent physical properties [14,50]. $BaZrO_3$ is a promising zirconate proton conductor widely used in refractory and electrical sectors. This material has excellent stability in a harsh environment, low proton migration, high melting temperature, high thermal expansion coefficient, excellent structure, and mechanical properties at high temperatures [51,52]. Furthermore, $BaZrO_3$ does not show any phase transition between low and high temperatures, making it suitable for electrochemical devices, including tritium monitoring systems, tritium recovery systems, hydrogen sensors, and hydrogen pumps [46,53,54].

Although cerate-based proton conductor like $BaCeO_3$ has the highest proton conductivity among other proton-conductor materials, it is unstable in water vapor and $CO_2$ atmosphere, whereas $BaZrO_3$ materials show stability in harsh weather (water vapor and $CO_2$) [16]. Alkaline earth zirconates, such as those found in $CaZrO_3$, $BaZrO_3$, and $SrZrO_3$, are typically more chemically stable and have more mechanical strength than alkaline earth cerate ceramics [55,56]. Many studies have shown that doping with $BaZrO_3$ can enhance proton conductivity and high chemical stability. The general formula of doping zirconate is $AZr_{1-x}D_xO_{3-\delta}$, where trivalent dopant D is used to replace the tetravalent Zr to create oxygen vacancy, which is crucial for proton-conduction perovskite ($ABO_3$) lattice structure [57]. The proton conductivity of the $BaZrO_3$ is greatly affected by the type and amount of the dopant used in the barium zirconate. With increasing Zr materials, the electrolyte sintering temperature is also increased, and as a result, the ionic conductivity is decreased [58]. Moreover, $BaZrO_3$ has high grain-boundary resistance which hinders electrochemical applications. Therefore, to improve the proton conductivity, it is essential to maintain a minimum grain-boundary resistance and high sintering temperature [59,60].

Studies have shown that Y-doped $BaZrO_3$ ($BaZr_{1-x}Y_xO_{3-\delta}$) exhibits excellent chemical stability with high proton conductivity [61]. For example, Liu et al. investigated $BaZr_{1-x}Y_xO_{3-\delta}$ electrolyte by partially replacing $Zr^{4+}$ with neodymium ($Nd^{3+}$) to enhance the sinterability and conductivity of the electrolyte [62]. The results showed that $BaZr_{0.7} Nd_{0.1}Y_xO_{3-\delta}$ had higher proton conductivity than $BaZr_{1-x}Y_xO_{3-\delta}$ electrolyte and that Nd³⁺ doping increased the chemical stability. However, neodymium (Nb) is a rare-earth element and expensive, which is not feasible for commercial application. On the other hand, mixed BaCeO₃-BaZrO₃ with dopant shows higher chemical stability but enriched Zr, restricting applications due to poor sintering and high grain-boundary resistance [63,64]. Therefore, further modification is required in zirconate to improve its proton conductivity with suitable stability for electrochemical application.

## 3. Electrochemical Hydrogen Device

Electrochemical devices are an essential scientific innovation enabling the development of an electric vehicle for the future. The principles of electrochemistry have materialized in hydrogen storage [65], hydrogen sensor [66], and hydrogen compressor [67] applications, as well as different chemical sensor applications. The basic electrochemical hydrogen devices have the following components: anode (electrode), electrolyte (proton-conducting solid), and cathode (electrode) (Figure 2) [68]. Electrochemical hydrogen devices use two fundamental principles: electromotive force (EMF) and the hydrogen transport phenomenon of the electrolyte. Recently, electrochemical devices have extensively used proton-conducting zirconates [16]. The small radius of protons enables the ions to fit into the interlayer structure of the cathode.

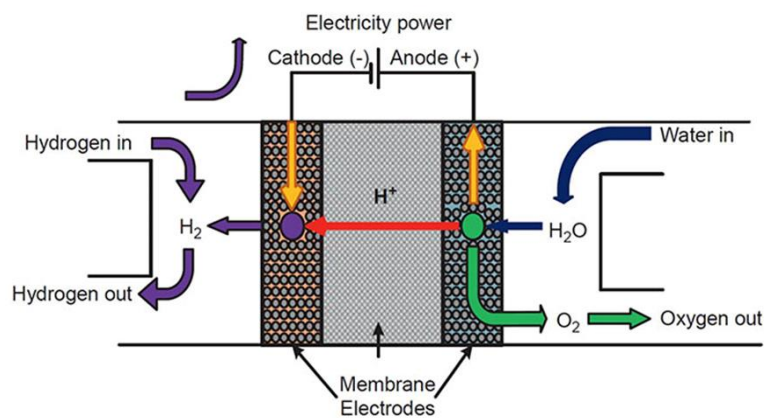

**Figure 2.** The fundamental design and operation of a proton-exchange membrane (PEM)-based electrochemical hydrogen device. Reproduced from Ref. [68]. Copyright © 2019 Elsevier.

These electrochemical devices utilize EMF the same as the principle of galvanic cells. The device is called a hydrogen sensor when EMF is used to produce signals. On the other hand, if the EMF force of the electrochemical cell is used to separate hydrogen, it is called a hydrogen pump. Radioactive isotopes like tritium (³H) can be separated using the same principles. An electrochemical reactor is necessary to convert water vapor and methane to tritium. Similarly, tritium can be monitored as a function of applied current, thus making this electrochemical device a platform for tritium monitoring. Moreover, separating radioactive molecules like Rn and enrichment of tritium can be effective for lower levels of tritium detection [69].

### 3.1. Tritium Monitoring Systems

From the public perception and safety-management perspectives of a fusion test facility, tritium monitoring faces unique and hazardous challenges. In that sense, detecting low levels of tritium has been one of the critical issues for the facilities that handle tritium. Tritium monitoring fulfills the purposes of (a) removal of interference of other radioactive materials, (b) discriminating between tritiated water vapor (HTO) and tritium molecules (HT), and (c) enrichment of tritium [70]. Usually, tritium monitors are constructed by combining an ionization chamber with semipermeable polymer membranes, such as silicone, Nafion, perfluorosulfonic acid resin, amide, etc. [71], which is used to selectively remove radioactive interference like radon and other chemical forms of tritium. The application of an electrochemical hydrogen pump using a proton-conducting oxide has also been proposed as a membrane separator [69].

Proton-conducting ceramics have unique characteristics that enable them to extract hydrogen from water vapor when DC current is passed through them at higher temperatures [71]. Among the proton-conducting ceramics, zirconates have shown the most promise because of their capability of higher hydrogen extraction. However, even within zirconates, $CaZr_{0.9}In_{0.1}O_{3-\alpha}$ and $SrZr_{0.9}Yb_{0.1}O_{3-\alpha}$ have shown the most potential as a tritium monitoring device, and their feasibility has been tested under various atmospheres and temperatures [72].

### 3.2. Tritium Recovery System

In a nuclear power system, especially in fusion reactors, tritium and its related compounds must be recovered for efficient recycling and radiation safety [73–78]. Typically in a fusion reactor, the radioactive isotopes are burned in high-temperature plasma, purged in an inert gas environment, recovered in a trap, and stored in a metal particle bed before being injected into the chamber again [79]. However, in more recent times, electrochemical cells have revolutionized the tritium fuel cycle and tritium recovery system [79]. The driving force of tritium transportation is enabled by electrochemical hydrogen (and related isotopes) transportation, whereby appropriate conductors can create the correct potential difference, which in turn can extract tritium selectively from a mixture of radiation gases [80]. Proton-conducting ceramic cells, mostly zirconates, have great potential to convert tritium (and related molecules) to electricity and vice versa so that they could be used for tritium (and other hydrogen isotopes) extraction [79]. Among the zirconate proton conductors, calcium and strontium zirconates are most widely used for their high conductivity and superior transfer properties [79,81]. Calcium zirconate with indium doping ($CaZr_{0.9}In_{0.1}O_{3-\alpha}$) is used in electrochemical tritium extraction cells [82], whereas strontium zirconate with ytterbium doping ($SrZr_{0.9}Yb_{0.1}O_{3-\alpha}$) is used in porous ($Ni/SiO_2$ and $NiO/SiO_2$) electrochemical cells at 600–700°C for tritium recovery [79].

### 3.3. Hydrogen Sensors

Hydrogen is one of the promising alternatives to fossil fuels because of its high efficiency and positive environmental impact. The rapid expansion of hydrogen in the energy market is becoming critically important for accurately sensing hydrogen in all its forms. Monitoring hydrogen can help prevent the formation of explosive/flammable/combustible mixtures. Other hydrogen sensors are used in various metallurgical processes and industrial chemical refineries [83]. Standard hydrogen sensors can detect the presence of hydrogen gas at the parts per million (ppm) level [84]. Even though electrochemical, electrothermal, optical, and even acoustic sensors are available in the market, electrochemical sensors are the most affordable, portable, and temporally stable [16].

Interestingly, proton-conductor ceramic materials have been used as electrolyte materials in electrochemical hydrogen sensor cells. Because of the ease of synthesis and flexibility in the temperature range, zirconate proton-conducting ceramics have been most widely used [85]. Earlier studies have reported that calcium-based zirconates ($CaZrO_3$) doped with indium (In), manganese (Mg), and scandium (Sc) can enhance physical and chemical properties, including proton conductivity and chemical stability [85–87]. These electrolytes can be used in conjunction with other sensing electrodes [83] and can act as self-reference electrodes [88]. These hydrogen sensors exhibit suitable sensing in various temperature ranges [89] and have superior reproducibility and stability [83].

### 3.4. Hydrogen Pumps

Hydrogen pumps are an essential tool used exclusively in nuclear reactor facilities, usually in the tritium recovery process. Hydrogen pumps provide advantages like minimal toxic emissions, carbon-free sustainability, and a cheaper renewable energy source. Several methods have been employed to materialize a hydrogen pump, for example, cryogenic distillation and pressure swing adsorption [90]. However, the most straightforward path for a hydrogen pump is through electrochemical cells [16]. When a direct current is delivered to the proton-conducting electrolyte, hydrogen at the anode is ionized to form protons in the electrolyte [16]. The electrolyte layer transforms the proton into hydrogen gas as it passes toward the cathode (Figure 2). This approach utilizes separated hydrogen in a regulated manner by applying an electrolyte current.

Proton-conducting perovskite oxides such as barium zirconate ($BaZr_{0.1}Ce_{0.7}Y_{0.2}O_{3-\delta}$), calcium zirconate ($CaZr_{0.96}In_{0.04}O_3$), and strontium zirconate ($SrZr_{0.9}Y_{0.1}O_{3-\alpha}$) are used as an electrolyte layer, and hydrogen is produced in the cathode [90–92]. Even though cerate-based perovskites can also be used as an electrolyte, cerate easily reacts with $H_2O$ and $CO_2$ (usually present in the by-product gases) [92]. Because of the chemical stability and relative reactivity against carbon dioxide, zirconate-based perovskites like doped strontium zirconate with yttrium have been used an electrolyte with palladium electrodes and utilized as a hydrogen pump [92]. To a similar end, scientists have used ytterbium doping for a similar hydrogen pump mechanism with $Ni/SiO_2$ porous electrodes [79]. Zirconate perovskites also have higher mechanical strength than their cerate counterparts, which gives durability in harsh nuclear fusion environments.

## 4. Electrochemical Device Applications

The recent progress of zirconate-based materials as an electrolyte in electrochemical device applications, namely, tritium monitoring, tritium recovery, hydrogen sensor, and hydrogen pump systems, their performance and improved properties are discussed and presented in the following section.

### 4.1. Tritium Monitoring Systems

Due to their potential radioactive properties, low-level tritium detection is an imperative issue in tritium handling facilities. Recently, the development and utilization of various types of tritium monitors to control the background count of radiation, detection between HTO and HT, and tritium enrichment have been reported. Tanaka et al. (2004) [70] studied the extraction of hydrogen using argon with 20.7% oxygen content as fed gas to the anode under a wet air environment. The purpose of using this type of gas composition was to reproduce the operating environment of the tritium stack monitor. In addition, pasted platinum electrodes have been used, as the hydrogen evolution rate decreased under a wet oxygen environment for plated platinum electrodes [81]. The performance of the hydrogen pump was evaluated in terms of temperature, water vapor partial pressure, and gas flow rate. First, fed gas was supplied to the anode at 600, 700, and 800 °C, 3.5 V, and 91 mL/min. It was observed that the hydrogen evolution rate did not start before 700 °C. However, the maximum evolution rate was found to be 0.64 mL/min at 800 °C, and the rate of hydrogen recovery was 58%. Another parameter, e.g., proton transport number, which refers to the ratio of protonic current to total current, has been reduced to 0.52 due to electron-hole migration through an electrolyte with the proton. Under the above conditions, the hydrogen evolution rate was observed as 0.65 mL/min and 0.52 mL/min for a gas flow rate of 137 mL/min and 47 mL/min, respectively. However, the hydrogen recovery rate was much higher for 47 mL/min, e.g., 96%, whereas for 137 mL/min, it was only 38%. The dew point of water vapor was reduced to around 0.042%. Furthermore, the hydrogen evolution rate depends on water vapor partial pressure. It was reported that the hydrogen evolution rate and recovery rate were 0.67 mL/min and 60%, respectively, for 1.2% water vapor. The evolution rate increased by 1.15 times when water vapor partial pressure changed from 0.86 to 1.2% at 91 mL/min. As such, it is imperative to consider the fluctuation of the hydrogen evolution rate during the design of a tritium monitor using proton-conducting oxide. Figure 3 shows a typical schematic of a tritium monitor with proton-conducting material [70].

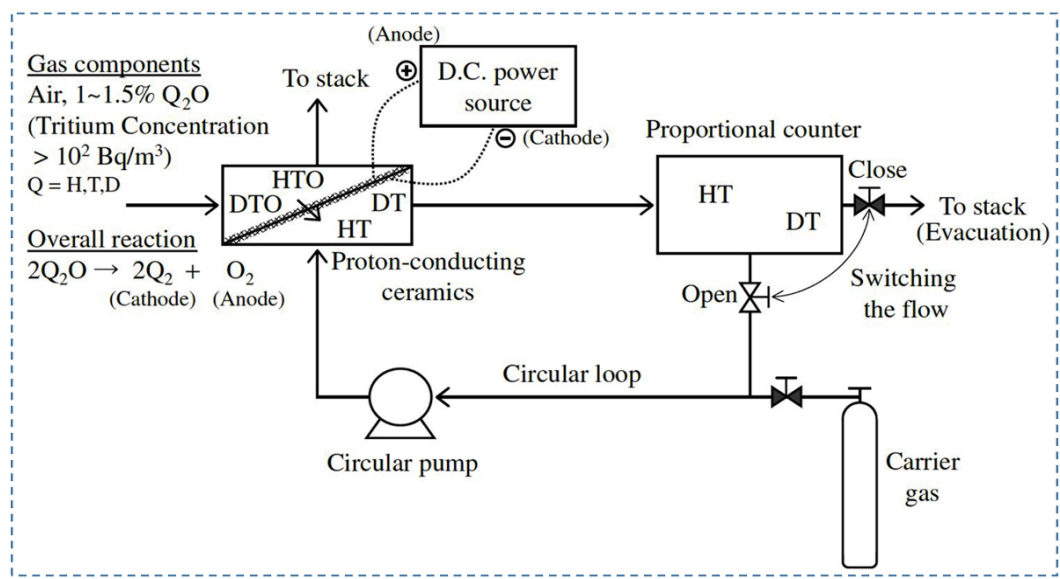

**Figure 3.** A schematic representation of the tritium monitor in combination with the proton-conducting material. Reproduced from Ref. [70]. Copyright © 2004 Taylor & Francis.

In another study, Tanaka et al. (2005) [72] investigated the effect of electrode attachment procedures on hydrogen extraction from exhaust gas mixture (methane, hydrogen, and water vapor) by using $CaZr_{0.9}In_{0.1}O_{3-\alpha}/SrZr_{0.9}Yb_{0.1}O_{3-\alpha}$. The plated platinum electrode (4 µm, 1 µm thickness) and pasted platinum electrode were used to evaluate the performance of the hydrogen pump. To simulate the exhaust gas composition, argon gas containing 0.1% $H_2$, 0.1% $CH_4$, and 1.2% $H_2O$-vapor was used as fed gas at 2 V. It was found that, at 800 °C, the evolution rate of hydrogen was 0.16 mL/min [93] for pasted electrodes, whereas for plated electrodes (4 µm), it was 0.46 mL/min, and 0.74 mL/min by $CaZr_{0.9}In_{0.1}O_{3-\alpha}$ and $SrZr_{0.9}Yb_{0.1}O_{3-\alpha}$, respectively. The lower hydrogen evolution rate for pasted electrodes was attributed to methane oxidation, where the water vapor's electrolysis produced oxygen, but for plated electrodes, the decomposition of methane and water vapor enhanced the hydrogen concentration (to 0.5%) on the anode side, leading to a higher evolution rate of hydrogen. On the other hand, in water vapor electrolysis with $SrZr_{0.9}Yb_{0.1}O_{3-\alpha}$ electrolytes under a wet argon environment, the total current increased, but the hydrogen evolution rate was not increased for plated platinum electrodes (4 µm). The flow of electron holes through the electrolyte along with proton was the reason for this phenomenon. Porous pasted electrode and plated electrode (1 µm) may reduce electron-hole flow and increase proton flow by providing a better diffusion path for oxygen from the electrode–electrolyte interface to the outside of electrodes. Therefore, the pasted platinum electrode is useful for $H_2O$-vapor, whereas the plated platinum electrode is useful for $H_2$ and $CH_4$.

Later, Tanaka et al. (2006) [71] studied the effect of a combination of a pasted platinum electrode as an anode and a plated platinum electrode wrapped with gold mesh as a cathode with $CaZr_{0.9}In_{0.1}O_{3-\alpha}$ electrolyte in a hydrogen pump. During performance testing, wet argon gas (1.2% water vapor) with 20% oxygen was supplied to the anode, and dry argon gas to the cathode at 100 mL/min, 923 K, and 2 V. In these conditions, the maximum evolution rate of hydrogen was observed due to the flow of current through the whole surface of the cathode (Figure 4a) [71]. Afterward, hydrogen enrichment in a closed-loop system was executed using this setup. Different percentages of water vapor concentration—0.7%, 0.86%, 1.2%, and 1.7%—were tested (Figure 4b) [71]. With the increase in water vapor concentration, the current decreased slightly, enriched hydrogen concentration increased, and the hydrogen evolution rate decreased with time. The enhancement in hydrogen concentration can be depicted by the increase in proton conductivity, which eventually decreases total conductivity. The reduction in the $H_2$-evolution rate was demonstrated by the enhancement of electrochemical hydrogen potential on the cathode side, so the performance of the hydrogen pump was not improved by hydrogen concentration enrichment below 1.7% in the closed-loop system. It was also reported that for extracting full water vapor under the described experimental conditions, the required electrode is 75 cm².

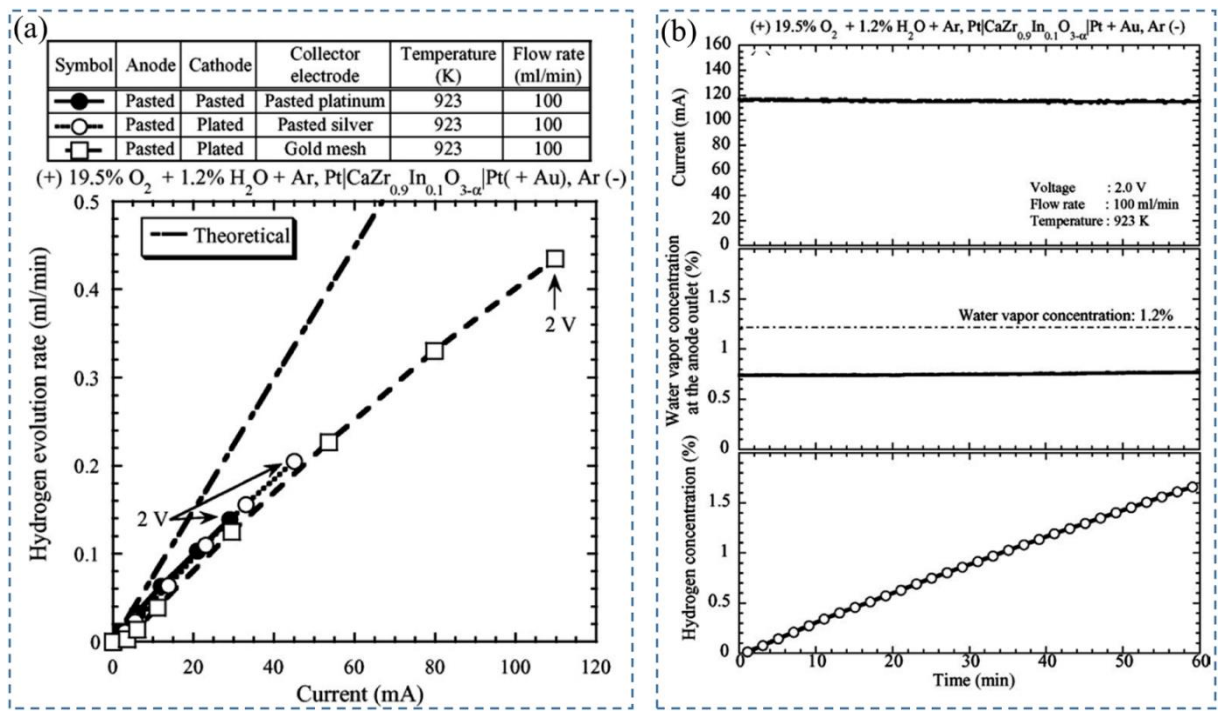

**Figure 4.** (**a**) Change in hydrogen evolution rate with time for different electrode combinations, (**b**)Time series response of current (top panel), water vapor concentration at the anode outlet (middle panel), and hydrogen concentration (bottom panel) against enrichment characteristics. Reproduced from Ref. [71]. Copyright © 2006 Elsevier.

Tanaka et al. (2008) [94] also designed a tritium monitor prototype with a closed-loop system consisting of $CaZr_{0.9}In_{0.1}O_{3-\alpha}$ as an electrochemical hydrogen pump. The performance of the hydrogen pump was assessed at a temperature range of 873 to 1073 K by water vapor electrolysis. It was found that the proton transport number decreased at high temperatures, so for extracting hydrogen, a high temperature was not required. In addition, to extract a large amount of hydrogen, a large current was required to pass with a thin specimen consisting of a large surface area while lower voltage and temperature can be maintained to protect the cell from being damaged. By satisfying these conditions, at 973 K, 0.5 A, and 0.5 mm thickness, the rate of hydrogen extraction was reported as 1.8 cm³/min. Further increase in current caused a reduction in proton transport numbers, which can be attributed to the reduction in $H_2O$-vapor partial pressure, rise in oxygen partial pressure in the anode, and an electron hole in a defect equilibrium reaction. Voltage decreased with increasing current. Furthermore, in the closed-loop system, the concentration of hydrogen was reported as 6% after 50 min at 0.625 A, 973 K.

To improve the performance of the hydrogen pump, Tanaka et al. (2008) [69] introduced a proportional countertype tritium monitor consisting of $CaZr_{0.9}In_{0.1}O_{3-\alpha}$ using HTO vapor to extract HT. After the extraction, HT was carried out by a metal bellow pump, followed by mixing with air and p10 gas. Then, the pulse count of the tritium monitor was recorded and converted to the tritium concentration. Experimental conditions were 973 K, electrode area 62 cm², feed gas flow rate 300 cm³/min, and 10 cm³/min for anode and cathode, respectively. The extraction was performed and controlled by an electric current, and the evolution was reported in terms of current density. It was reported that tritium concentration increased with current density. Nevertheless, at a constant current density, tritium concentration increased, whereas the hydrogen recovery rate decreased with increasing water vapor partial pressure on the anode side (Figure 5a) [69]. This phenomenon can be demonstrated by the inverse relationship of the hydrogen recovery rate with water vapor partial pressure [95] In addition, the measured concentration of the tritium using the proposed tritium monitor was 15% higher than the introduced concentration. Figure 5b shows the relationship between the tritium concentration in the compartmented anode and the estimated tritium concentration [69].

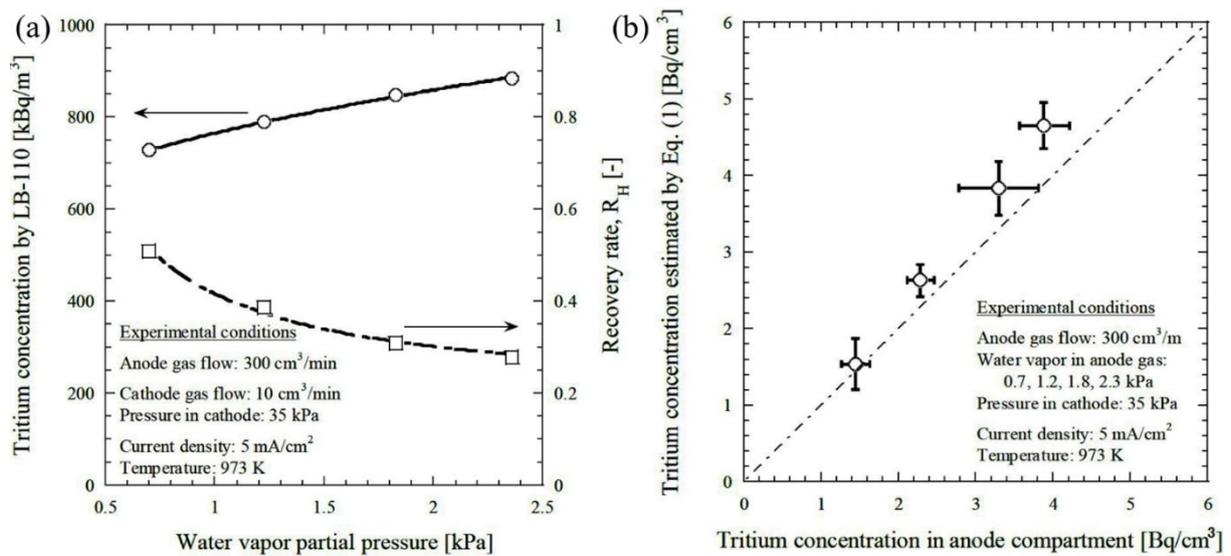

**Figure 5.** (**a**) Change in tritium concentration and hydrogen recovery rate with water vapor partial pressure, (**b**) relationship between the tritium concentration in anode compartment the estimated tritium concentration. Reproduced from Ref. [69]. Copyright © 2015 Taylor & Francis.

Tanaka et al. (2011) [43] designed a tritium monitor to extract the hydrogen and tritium over 873 K to 1073 K. In this process, the electrolysis of HTO was achieved by a hydrogen isotope pump that used $CaZr_{0.91}In_{0.1}O_{3-\alpha}$ as a membrane separator. It was observed that the maximum evolution rate of both hydrogen and tritium was at 973 K. The decreasing solubility of proton and electrode reaction's kinetic demonstrated a lower evolution rate at a higher temperature (>973 K). In the hydrogen isotope pump, the effect of isotope was found, which was attributed to the lower T/H ratio on the cathode side than the anode side, while on the anode side the T/H ratio was extremely low. This low T/H ratio also confirmed the possibility of extracting the tritium gas with hydrogen. In order to reduce the isotope effect, it was suggested by the research group to use a large electrode area so that the applied current can be increased, which eventually leads the hydrogen recovery rate to unity. Table 1 shows zirconate-based tritium monitors along with their relevant parameters.

**Table 1.** Zirconate-based tritium monitoring systems.

| Electrolyte Materials | Proportional Counter Volume (m³) | Mixing Ratio | Flow Rate (L/min) | Sensitive Volume (L) | Tritium Counting Efficiency (%) | Calibration Factor (kBq/m³ per cps) | Background Rate in Tritium Channel (cps) | Minimum Detectable Activities (Bq/m³) | Ref |
|---|---|---|---|---|---|---|---|---|---|
| $CaZr_{0.9}In_{0.1}O_{3-\alpha}$ | 0.0013 | 1:04 | 0.2 air/0.8 P10 | 0.26 | 55 | 7 | 0.4–3 | 100–5400 | [69] |
| $CaZr_{0.9}In_{0.1}O_{3-\alpha}$ | 0.0013 | N/A | 0.091 | N/A | N/A | N/A | N/A | 1300 | [70] |

## 4.2. Tritium Recovery Systems

Tritium recovery from exhaust gas is an important issue for nuclear fusion reactors. For this purpose, the palladium membrane is commercially used as a diffuser, but researchers are looking forward to an alternative due to its high cost and maintenance. Proton-conducting oxides showed promising characteristics in this regard. Tanaka et al. (2004) [93] demonstrated hydrogen extraction using a hydrogen pump consisting of a tube closed at one end made of a proton-conducting oxide as $CaZr_{0.9}In_{0.1}O_{3-\alpha}$ due to its superior stability in the chemical environment and mechanical reliability. It was found that for a gas flow rate of 91 mL/min and fed gas (to anode) with 0.1% hydrogen at dry or wet conditions, experimental results of hydrogen evolution rate showed good agreement with theoretical results for applied voltage 0–3.5 V, except 2.5 V. The evolution rate of hydrogen exceeded the maximum value below 2 V in wet conditions. This phenomenon was demonstrated by the complete extraction of supplied hydrogen followed by the electrolysis of water vapor on the anode side.

On the other hand, for fed gas with 1% $H_2$, the extraction of supplied hydrogen to the cathode side was 17% and 50% for the dry and wet conditions at 3.5 V, respectively. In addition, to investigate the water vapor electrolysis, 1.2% $H_2O$-vapor-carrying argon gas was supplied at different flow rates (47, 91, 137 mL/min) to the anode at 3.5 V. In these conditions, the maximum evolution rate of hydrogen was reported as 0.67 mL/min at 137 mL/min. However, the "current efficiency" decreased up to 80% with current due to the increase in the partial pressure of oxygen and a decrease in the partial pressure of water vapor on the anode side. The contribution of methane decomposition in the evolution of hydrogen in the dry condition with 0.1% $CH_4$ in argon gas was not even observed, whereas in the wet condition, the maximum rate of evolution was recorded as 0.37 mL/min while 3.5 V was applied. It was only 10% of evolved hydrogen on the cathode side. However, the maximum hydrogen evolution rate by $CaZr_{0.9}In_{0.1}O_{3-\alpha}$ from the gas mixture (0.1% $H_2$, 0.1% $CH_4$, 1.2% $H_2O$) was found at 0.34 mL/min under the wet condition and flow rate 91 mL/min. The decomposition of methane under wet conditions was initiated by the produced oxygen from the electrolysis of water vapor after the complete extraction of hydrogen in the anode for the pasted platinum electrode. Figure 6 shows the schematic of a typical tritium extraction system and the response of hydrogen evolution rate against various gases [93].

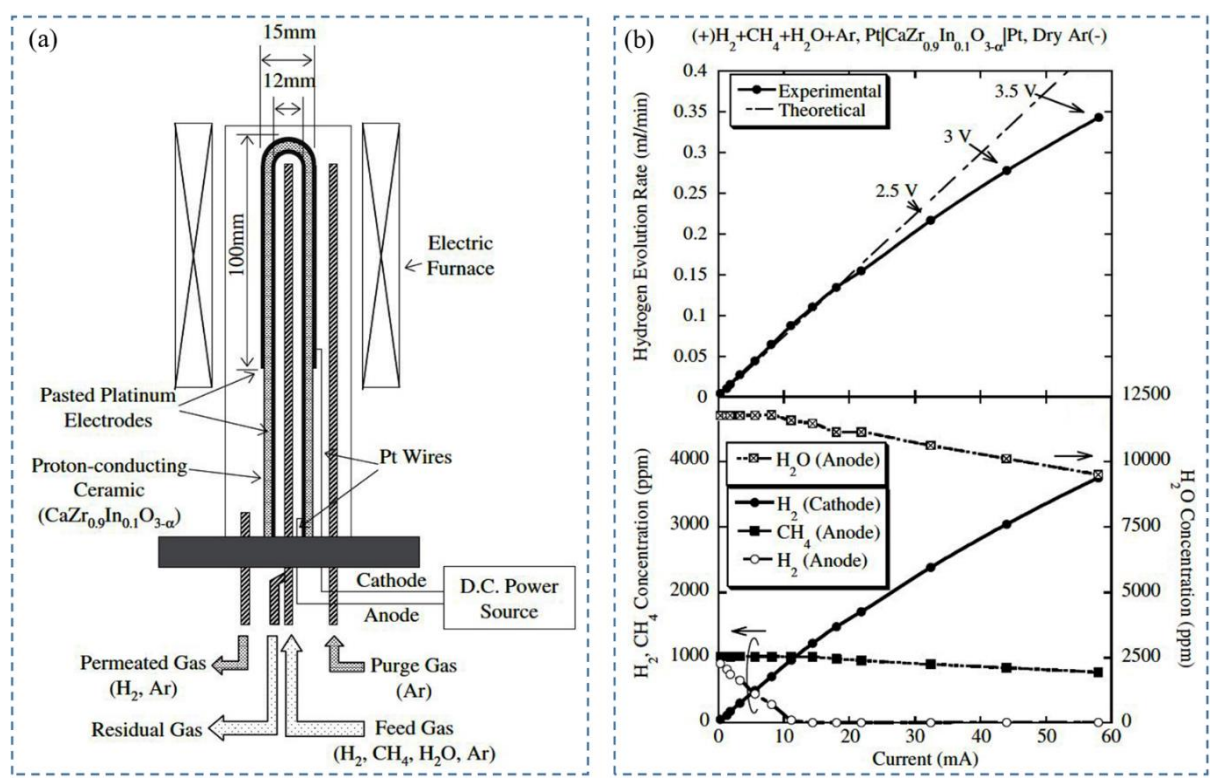

**Figure 6.** (**a**) A schematic representation of hydrogen extraction system with proton-conducting ceramic on one end, (**b**) response of hydrogen evolution rate (top panel) and hydrogen, methane, and water vapor concentration (bottom panel) as a function of current. Reproduced from Ref. [93]. Copyright © 2004 Taylor & Francis.

Later, to enhance the hydrogen extraction rate and electrode reactions for real-time application, Tanaka et al. (2004) adopted a new technique of preparing electrodes that includes electroless platinum plating on yttria-stabilized zirconia (YSZ) instead of using platinum paste [81]. For hydrogen pump and water vapor electrolysis experiments, wet argon gas with and without 1% hydrogen was used as fed gas into the anode. The applied voltage was between 2 and 0 V. SEM showed that the plated electrode surface consisted of small particles and pores of about 10 nm in size, while the pasted surface consisted of 1 μm pores (Figure 7) [80]. It was reported that the hydrogen evolution rate in the hydrogen pump increased by 3.2 times and the current increased by 3.5 times for plated electrodes compared to the pasted one. This phenomenon was attributed to the presence of a highly dense three-phase boundary due to the small particles. On the other hand, in the experiment of water vapor electrolysis, the current increased 1.5 times, but the hydrogen evolution rate decreased by 10%. With the presence of small pores on the plated surface, oxygen cannot diffuse easily into the electrode–electrolyte interface, which eventually increases the partial pressure of oxygen in the anode. As a result, electron-hole current increased, and "current efficiency" became less than unity, so the hydrogen evolution rate decreased.

In another study, Tanaka et al. (2012) [82] investigated the recovery of hydrogen isotope under vacuum on the cathode side using $CaZr_{0.9}In_{0.1}O_{3-\alpha}$ electrolyte with pasted and plated electrode. In addition, the mass transfer process was also evaluated. Experimental results showed that the performance of the hydrogen pumps improved under vacuum. It was demonstrated by reducing voltage between electrodes under vacuum conditions after applying direct current compared to the atmospheric pressure, irrespective of the electrode attachment procedures. Electrochemical impedance spectroscopy (EIS) was used to observe the mass transfer process. Two important impedance parameters, $W_{sr}$ and $W_{sc}$, confirmed that gas diffusivity increases relatively using pasted electrodes under vacuum conditions. Later, the microstructure study found that the larger pore size of pasted electrodes led to better gas diffusivity.

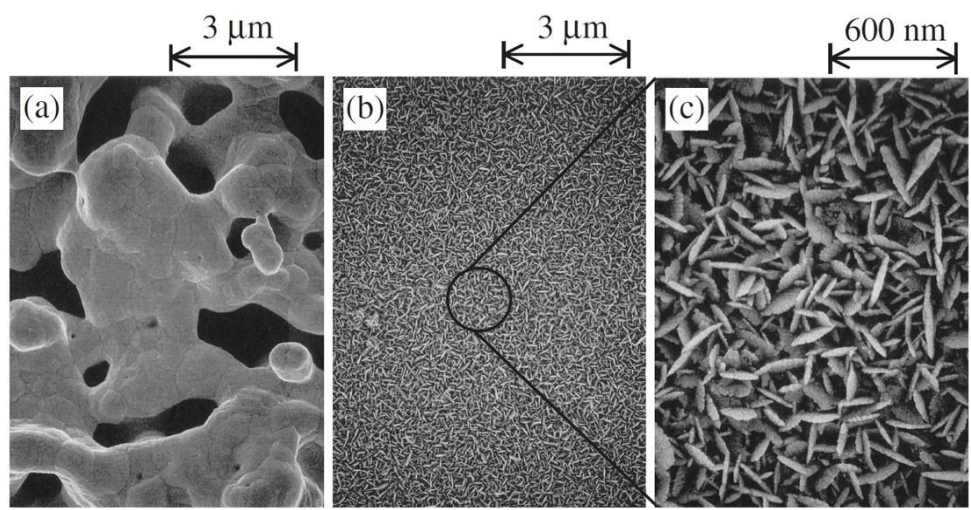

**Figure 7.** SEM of electrode surface; (**a**) pasted electrode (a magnification of ×10,000), (**b**) plated electrode (a magnification of ×10,000) and (**c**) plated electrode (a magnification of ×50,000). Reproduced from Ref. [81]. Copyright © 2004 Taylor & Francis.

Tirui et al. (2014) [80] also used tube-type $CaZr_{0.9}In_{0.1}O_{3-\alpha}$ to extract hydrogen efficiently and economically by an electrochemical hydrogen pump. According to Faraday's law, hydrogen extraction was performed from the mixture of He and $H_2$ at a voltage of less than 3.5 V DC. Various parameters, such as hydrogen concentration, temperature, current density, and voltage control hydrogen extraction, were used. It was found that higher hydrogen concentration enhanced the performance of the hydrogen pump. For example, at 2.35 V, 973 K, and 664 ppm, the hydrogen evolution rate was found as 0.0672 mL/min whereas for 2.61 V, 973 K, and 1128 ppm, it was 0.1046 mL/min, which indicates a 55% increase in hydrogen evolution rate. The performance of the hydrogen pump was characterized by the temperature range 873 K–1073 K, but the maximum hydrogen evolution rate was 0.12 mL/min at 1023 K, while the $H_2$ and current density were 0.12% and 0.113 mA/cm². Although the maximum $H_2$ evolution occurred at 1.34 V, it coincided with the theoretical value at 1.15 V (Figure 8a). Figure 8b shows the hydrogen evolution rate against the applied current [80]. In addition, under these conditions with the feed gas flow rate of 100 mL/min, the efficiency of hydrogen evolution of the hydrogen pump was recorded as higher than 99%, and the transportation number was around 1.0. Furthermore, the estimated electrolytic current and number of ceramic tubes were reported as $9.46 \times 10^{-3}$ A and 25, respectively, for the CIPITISE-TSE application at CIAE.

The effect of another proton-conducting oxide, $SrZr_{0.8}In_{0.2}O_{3-\alpha}$, on the performance of hydrogen pumps under various environments and temperatures (673–873 K) was investigated by Tanaka et al.

(2008) [96]. Wet argon was passed to the anode while pressure and current were 1.22 kPa and 75 mA, respectively. Results showed that with the increase in temperature, voltage and proton transport number (PTN) decreased under water vapor electrolysis. Maximum PTN was observed at 823 K and started to decrease with temperature. Similar results were found under these conditions when helium-carrying water vapor and hydrogen were passed to the anode at a pressure of 1.22 kPa and 1 kPa, respectively, but the transport number of oxygen ions increased with temperature. In both cases, mixed conduction occurred, which is responsible for the lower PTN. It was concluded that a higher temperature was not necessary to enhance hydrogen pump performance and a low temperature is practically favorable to reduce the migration of oxygen ions, leading to lower oxide degradation. The authors also studied the impact of oxygen partial pressure on the cathode side on hydrogen pump performance. The pressure was maintained over the range of 0.02 Pa to 1 kPa. It was found that the performance of the hydrogen pumps enhanced with increasing oxygen partial pressure, which can be attributed to the enhancement of proton transport numbers (above $10^2$ Pa) and reduction of voltage. Furthermore, the comparison of the performance of the hydrogen pump based on $SrZr_{0.8}In_{0.2}O_{3-\alpha}$ and $CaZr_{0.9}In_{0.1}O_{3-\alpha}$ was studied during underwater vapor electrolysis at 873 K. Even though both showed mixed conduction at elevated temperatures, proton conductivity and electrical resistance were lower for $SrZr_{0.8}In_{0.2}O_{3-\alpha}$ than $CaZr_{0.9}In_{0.1}O_{3-\alpha}$-based hydrogen pumps. These results indicate that proton conductivity is the major parameter to enhance the performance of the hydrogen pump.

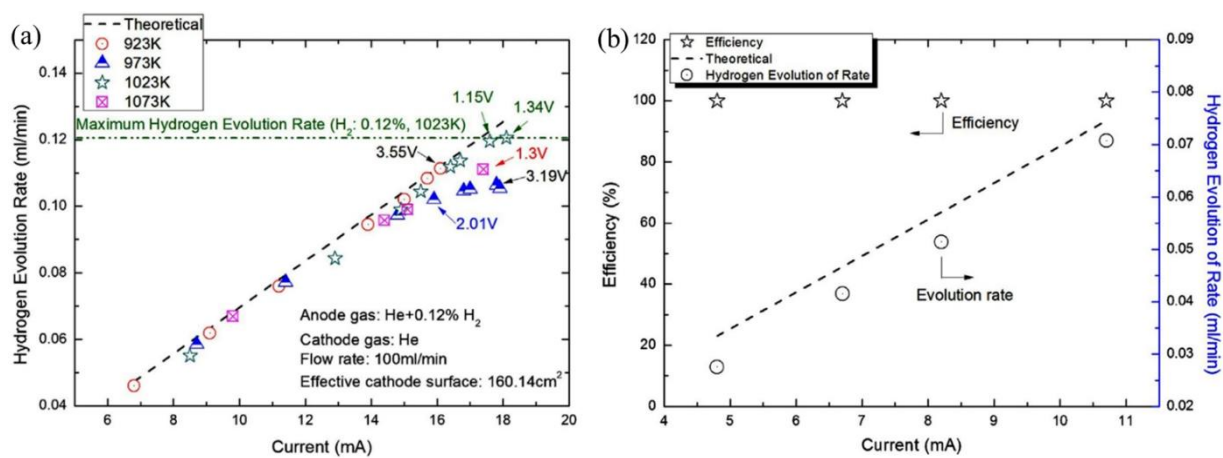

**Figure 8.** (**a**) Hydrogen evolution rate versus current under various temperatures, (**b**) The hydrogen evolution rate, and efficiency as a function of current (gas mass flow rates). Reproduced from Ref. [80]. Copyright © 2014 Elsevier.

Tanaka et al. (2010) [97] also used $SrZr_{0.9}Yb_{0.1}O_{3-\alpha}$ to extract hydrogen from a gas mixture. To simulate the gas mixture present in a hydrogen processing unit, argon gas carrying hydrogen and methane at 0.1 kPa and water vapor at 1.2 kPa were passed into the system at 973 K, 85 cm³/min. The hydrogen and water vapor concentration at the anode outlet was observed as 0.5% and 1%, respectively, while methane was not detected. This phenomenon was successfully demonstrated by the decomposition of methane and part of the water vapor into hydrogen gas. Considering the hydrogen pump composed of 0.5 kPa $H_2$ and 1 kPa water vapor, it was found that with the increasing current density (up to 1.1 mA/cm²) hydrogen evolution rate increased linearly along with the theoretical value. After 1.1 mA/cm², it increased nonlinearly, but became much smaller than the theoretical values, which depicted the migration of other charge carriers to the proton conductor. However, until 0.4 V, 1.1 mA/cm², hydrogen extraction to the cathode was executed by hydrogen pumping, while PTN was unity. Then, hydrogen extraction was executed by water vapor electrolysis, and the presence of oxygen was observed at 0.95 V. Later, oxygen supplied electron holes and oxide ions. These carriers are responsible for the reduced efficiency of the hydrogen pump and the reduction of electrolytes. This issue could be solved by reducing the operating temperature to less than 773 K.

Fukada et al. (2007) [79] conducted a study that proposed a ceramic cell with one end closed tube consisting of $SrZr_{0.9}Yb_{0.1}O_{3-\alpha}$ to recover tritium from the exhaust gas. The porous electrodes of this cell were composed of Ni/SiO$_2$ and NiO/SiO$_2$. The fed gas supplied to the anode was wet methane or hydrogen, whereas for the cathode, it was wet oxygen. Results showed that regardless of the supplied conditions (CH$_4$ + H$_2$O, H$_2$ + H$_2$O), all the *j-V* curves can be correlated by the linear equation $V = E_0 - jd/\sigma$ under direct-current (DC) method or a temperature range 600–700 °C. In addition, EMF ($E_0$) and overall conductivity ($\sigma$) were determined in terms of the partial pressure of H$_2$O on the anode side and the temperature. It was observed that the mass transfer process of CH$_4$ + H$_2$O's products CO and H$_2$ took place by counter diffusion and normal diffusion method, respectively, through the porous anode, which eventually affected $E_0$. Conversely, overall proton conductivity was affected by the reduction reaction rate of oxygen between the interface of ceramic and the cathode NiO/SiO$_2$. This phenomenon was similar for both supplied conditions. Additional tests were conducted by feeding CH$_4$ without H$_2$O into the fuel-cell system. From SEM-EDX, it was found that carbon deposition occurred at the cathode interface, which was not observed with H$_2$O until six months. All these findings depicted the stable operation of SrZr$_{0.9}$Yb$_{0.1}$O$_{3-\alpha}$ ceramic cell to extract Q$_2$ (Q = H, D, T) successfully. Table 2 shows different zirconate-based tritium recovery systems and their related parameters.

**Table 2.** Zirconate-based tritium recovery systems.

| Electrolyte Materials | Sintering Condition | Sample Gas Mixture Used at Anode | Sample Gas Mixture Used at Cathode | Electrode Type | Operating Temperature (°C) | Ref. |
|---|---|---|---|---|---|---|
| CaZr$_{0.9}$In$_{0.1}$O$_{3-\alpha}$ | 800 °C for 1 h | He + H$_2$ | He | Pt | 650–800 | [80] |
| SrZr$_{0.9}$Yb$_{0.1}$O$_{3-\alpha}$ | N/A | CH$_4$ + H$_2$ | O$_2$ + H$_2$O | Ni/SiO$_2$ and NiO/SiO$_2$ | 600–700 | [79] |
| CaZr$_{0.9}$In$_{0.1}$O$_{3-\alpha}$ | N/A | N/A | N/A | Pt | 800 | [93] |
| CaZr$_{0.9}$In$_{0.1}$O$_{3-\alpha}$ | N/A | N/A | N/A | Pt | 800 | [81] |
| CaZr$_{0.9}$In$_{0.1}$O$_{3-\alpha}$ | N/A | N/A | N/A | Pt | 700 | [82] |

### 4.3. Sensor Devices

Perovskite materials such as ACeO$_3$ (where A is either Ba or Sr) can act as a high proton-conducting electrolyte in hydrogen sensor devices, but their poor chemical stability limits their application [98,99]. However, indium-doped CaZrO$_3$ shows better chemical stability than ACeO$_3$ at higher operating temperatures, but has relatively lower proton conductivity [100]. Moreover, the sensing properties of the sensor devices mainly depend on the proton conductivity, and conductivity depends on the doping effect and phase composition. Therefore, the solid-state reaction method produces the In-doped CaZrO$_3$ to improve the sensitivity. Chen et al. (2016) [89] investigated the connection between the calcination temperature and the fabrication of CaZr$_{0.9}$In$_{0.1}$O$_{3-\delta}$ electrolyte during the solid-state reaction procedure. Figure 9 shows the schematic of the hydrogen sensor and the experiment setup [89]. It was found that CaZrO$_3$ can be produced from ZrO$_2$ and CaO directly at 1000 °C or at two stages creating an intermediate product (CaZr$_4$O$_9$) at a higher temperature of 1000 °C to 1200 °C. The indium-doping process mainly occurs in CaZrO$_3$ between 1200 °C and 1400 °C. From the *Emf* response of the sensor, the authors found out that the Emf value of the sensor based on CaZr$_{0.9}$In$_{0.1}$O$_{3-\delta}$ electrolyte is close to the theoretical value and has a proton transport number close to unity (0.95). This proves the excellent sensing capability of the indium-doped electrolyte. On the other hand, due to the lower proton transport number (0.85) and Emf, undoped CaZrO$_3$ does not show good sensing properties.

Matsumoto and Iwahara (2000) [101] developed an electrochemical device that can be used to detect hydrogen and deuterium by using EMFs of cells with the proton-conducting electrolyte CaZr$_{0.90}$In$_{0.10}$O$_{3-\alpha}$. The authors proposed that the EMF of the cell had two components: (1) voltage difference due to the difference in mobility of deuterium and proton, and (2) different reaction rates of D$_2$ and H$_2$ at the platinum electrodes. The EMF measured at 700 °C of the H$_2$ | | D$_2$ cell was found to be 16.9 mV which was constant in all atmospheres and similar to a previous study by the same authors [102]. The cell's EMFs linearly increased with deuterium molar concentration in the H$_2$ and D$_2$ gas mixture. The fabricated hydrogen isotope sensor was very stable and responsive.

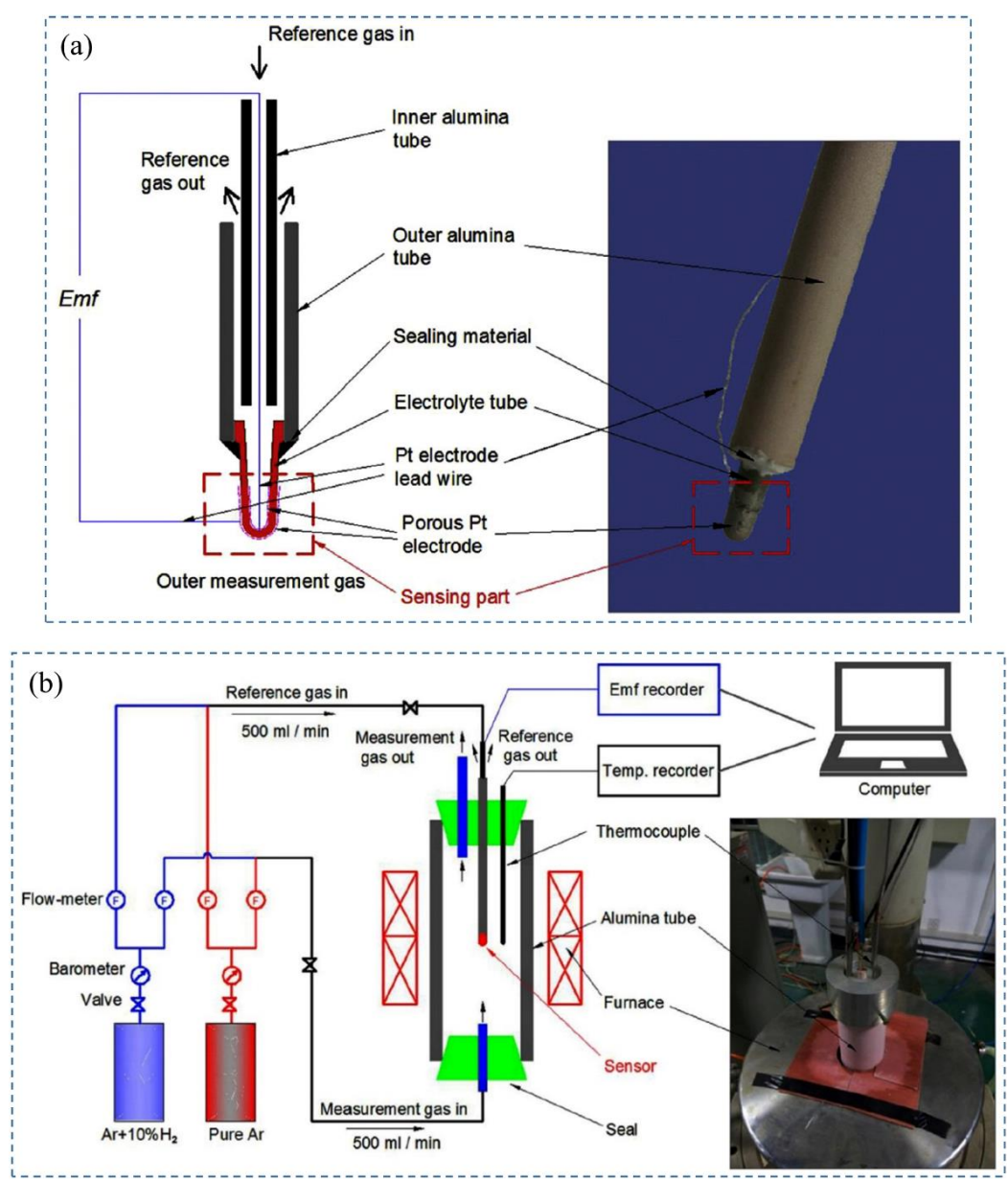

**Figure 9.** (**a**) Schematic of $H_2$ sensor with a sintered $CaZr_{0.9}In_{0.1}O_{3-\delta}$ (**b**) Experimental setup for hydrogen sensor in various partial pressure gases. Reproduced from Ref. [89]. Copyright © 2016 Elsevier.

Kalyakin et al. (2019) [103] studied amperometric-type hydrogen sensors with $CaZr_{0.95}Sc_{0.05}O_{3-\delta}$ electrolyte in the application of humidity analysis. The sensor consists of two electrodes: YSZ having oxygen-ionic conductivity and $CaZr_{0.95}Sc_{0.05}O_{3-\delta}$ having protonic and oxygen-ionic conductivity, produced by glycine glycerin nitrate combustion synthesis. The sensor was tested in wet air, with a partial pressure of vapor ranging from 0.025 atm to 0.11 atm. From the voltage-current analysis, it was seen that with the increase of voltage across the electrode of the sensor, the current rises to a limiting value, and the higher the partial pressure is, the higher the limiting current. XRD pattern and SEM showed that $CaZr_{0.95}Sc_{0.05}O_{3-\delta}$ crystal has an orthorhombic perovskite structure, and when sintered at 1550 °C, it has well-formed and dense grains with a relative density of 99%, which contributes to its good sensing properties. From the conductivity analysis, the authors noticed that $CaZr_{0.95}Sc_{0.05}O_{3-\delta}$ electrolyte shows electronic and ionic conductivity at high temperatures (800 °C), but only ionic conductivity at a lower temperature (600 °C and 700 °C) due to the constant conductivity value over the range of different partial pressure of oxidation atmosphere [103]. It can be concluded from their study that $CaZr_{0.95}Sc_{0.05}O_{3-}$ $_\delta$ can be used as a mix-ionic conductive electrolyte at relatively high temperatures and in dry oxidation atmospheres.

The performance of amperometric hydrogen sensors primarily depends on the electrolytes and sensing electrodes. Among the proton-conducting electrolytes, $In^{3+}$-doped calcium zirconate is chemically stable and has been used in hydrogen sensors [104,105]. On the other hand, the performance of nanostructured ZnO as sensing electrodes has been studied recently [106–109]. Dai et al. (2012) [83] have proposed an amperometric hydrogen sensor with a bilayer proton conducting electrolyte, $CaZr_{0.9}In_{0.1}O_{3-\alpha}$ and ZnO electrode as sensing materials. Figure 10 shows the fabrication process of the proposed hydrogen sensor [83]. The principle of the sensors is based on current measurement in the case of applied voltage on the nanostructured ZnO anode. From the XRD pattern, it can be concluded that $CaZr_{0.9}In_{0.1}O_{3-\alpha}$ is stable under pure $CO_2$, $H_2$, and wet Ar atmospheres. Arrhenius's plot of the proton conductivities of the electrolyte shows that it has a good proton conductivity, making it an excellent choice for hydrogen sensors. After using ZnO as a sintering aid in the electrolyte, it is noticed from SEM images that ZnO effectively densifies the electrolyte. ZnO also helps to improve the sensor response by increasing the reaction sites due to its porous nature. I–V characteristics of the sensor exhibit the reversibility property, and the current response increase when the hydrogen concentration changes from 0 ppm to 500 ppm. The amperometric transient graphs show that the response time and recovery time increase with the increase in temperature because of the higher electrochemical kinetics at increased operating temperature. The response/recovery times are inversely proportional to the gas flow rate. Loading of the sensor materials also has an impact on the sensitivity of the sensor. ZnO 5 mg exhibits a higher current response compared to the 2 mg of ZnO loading. The hydrogen sensor is also stable under extended exposure to hydrogen gas at 700 °C. Further research is needed to improve this sensor's sensitivity and long-term stability.

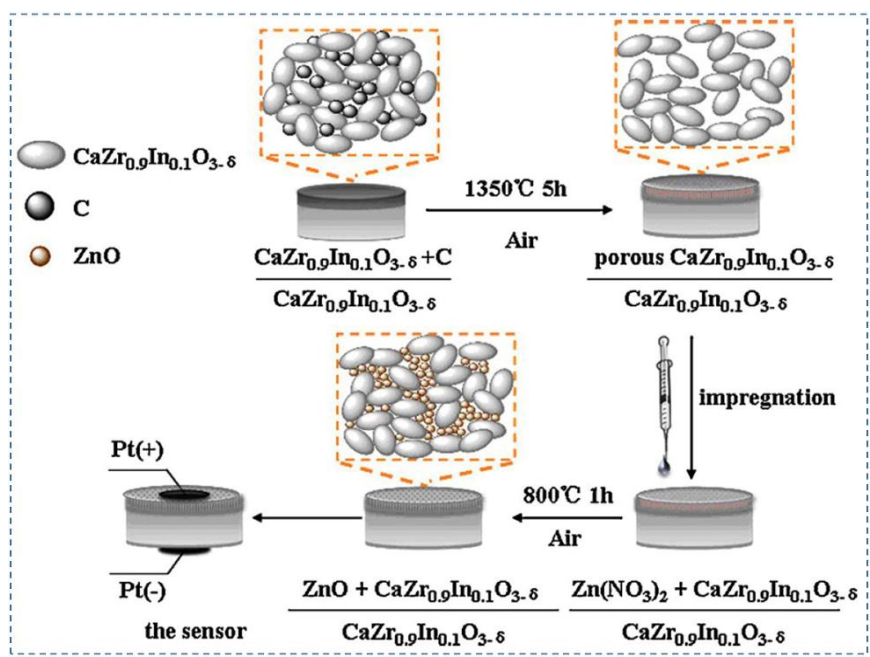

**Figure 10.** Chemical composition of materials and the schematic of the fabrication process of the hydrogen sensor. Reproduced from Ref. [83]. Copyright © 2012 Elsevier.

Ohshima et al. (2010) [110] investigated the transportation of hydrogen through molten salt LiF-NaF-KF (Flinak) by a hydrogen sensor consisting of proton-conducting oxide as a solid electrolyte. At first, the hydrogen sensor was validated by exposing it to an environment of Ar-$H_2$. In this environment, the partial pressure of hydrogen was known. During this test, it was observed that the hydrogen sensor showed stability in Flinak while the temperature was not changing. In addition, the EMF of the sensor showed agreement with the theoretical EMF at this condition. Nevertheless, with the temperature change, the EMF value varied significantly from the theoretical EMF. The temperature change was measured between the free surface and the position where the sensor was immersed. Hydrogen transport through Flinak was controlled by the temperature dependence of Henry's law. Its value was calculated by using the measured EMF. It was found that Henry's constant has negative temperature dependence for the transportation of hydrogen through Flinak.

Okuyama et al. (2016) [88] investigated the proton dissolution mechanism and electrochemical properties of a hydrogen sensor with proton-conducting Mn-doped calcium zirconate ($CaZr_{1-x}Mn_xO_{3-d}$, x = 0.05, 0.005) electrolyte. The conduction and protonation of the electrolyte were primarily analyzed when the electrolyte was exposed to a hydrogen atmosphere. According to the study of Kurita et al. (2019) [111], when the reference electrode has a low proton transport number, the EMF of the proton conductor mainly depends on the hydrogen activity of the working electrode. Therefore, developing a hydrogen sensor using $CaZr_{1-x}Mn_xO_{3-d}$ does not require knowing the hydrogen concentration of the reference gas. This allows the electrolyte to serve as a self-standard electrode. Experimental conditions include flowing wet air to the reference electrode and flowing different ratios of hydrogen and argon gas to the working electrode. Electron spin resonance (ESR) analysis suggests that hydrogen flow causes a reduction in manganese ions from $Mn^{3+}$ to $Mn^{2+}$. IR diffusion–reaction analysis, which is observed at 3300 cm$^{-1}$ and caused by O−H vibration, also shows similar spectra, suggesting that from the reduction of Mn ion, $CaZr_{1-x}Mn_xO_{3-d}$ acquires protons. Hence, conductivity increases with the increased hydrogen flow since the reduction of manganese ions increases proton conduction. The activation energies of conductivity for $CaZr_{0.995}Mn_{0.005}O_{3-d}$ and $CaZr_{0.95}Mn_{0.05}O_{3-d}$ in reducing atmosphere ($H_2/H_2O$) are 0.63 eV and 0.66 eV, respectively, whereas in an oxidizing atmosphere ($O_2/H_2O$), they are 1.23 eV and 1.10 eV, respectively. This result suggests that the predominant charge carrier differs in these two atmospheres. Moreover, from the EMFs of hydrogen concentration cells, it is concluded that $CaZr_{1-x}Mn_xO_{3-d}$ exhibits hole conduction in the oxidizing environment during proton conduction in a reducing environment. Table 3 shows the zirconate-based hydrogen sensors and the related parameters.

### 4.4. Hydrogen Pumps

Hydrogen energy is very environmentally friendly, as opposed to carbon-based fuel (oil, natural gas), which leads to global warming. An efficient method of hydrogen production is an important topic that was investigated by Sakai et al. 2008 [112]. Figure 11 shows the operating principle of the electrolysis mechanism, the SEM characterization of zirconate, and the hydrogen evolution rate for different electrodes [112]. For electrolysis, different kinds of electrode materials were used, e.g., platinum as both cathode and anode, $Sr_{0.5}Sm_{0.5}CoO_3$ (SSC-55) as an anode, Ni with an interlayer of $SrCe_{0.95}Yb_{0.05}O_{3-\alpha}$ (SCYb) as cathode, and $SrZr_{0.9}Y_{0.1}O_{3-\alpha}$ (SZY-91) with partial doping of cerium, i.e., $SrZr_{0.5}Ce_{0.4}Y_{0.1}O_{3-\alpha}$ (SZCY-541) as electrolyte. This type of electrochemical device is known as a hydrogen pump. The proton-conducting electrolyte used in their study was prepared by the solid-state reaction method. To overcome one of the limiting factors of the performance, i.e., overpotentials, the authors used SZY-91 electrolytes with different types of electrodes instead of platinum. When platinum electrodes with SZY-91 electrolyte were used for electrolysis, overpotentials were found to be very high, indicating electrolysis's poor efficiency. This is due to the electric current conduction and ionic currents during the electrolysis process. The authors were not sure whether this electric current is caused by holes or electrons, which makes room for further investigation in the future. When the SSC-55 anode and Ni cathode were used in place of platinum electrodes, the overpotential was significantly lower. SSC-55 | SZY-91 | Ni cell showed improved efficiency of current. As SCYb was inserted between the electrolyte and Ni cathode, the overpotential of the cathode was significantly reduced, but the hydrogen evolution rate significantly differed from the theoretical value in the high current density region. As such, SZCY-541 was used instead of the SZY-91 electrolyte to improve the current efficiency. In this case, overpotential was slightly smaller than using SZY-91, yet the current efficiency, as well as hydrogen evolution rate, was significantly improved. Therefore, the authors concluded that partial doping of Ce improved the cell's performance by reducing electric current and enhancing ionic transport.

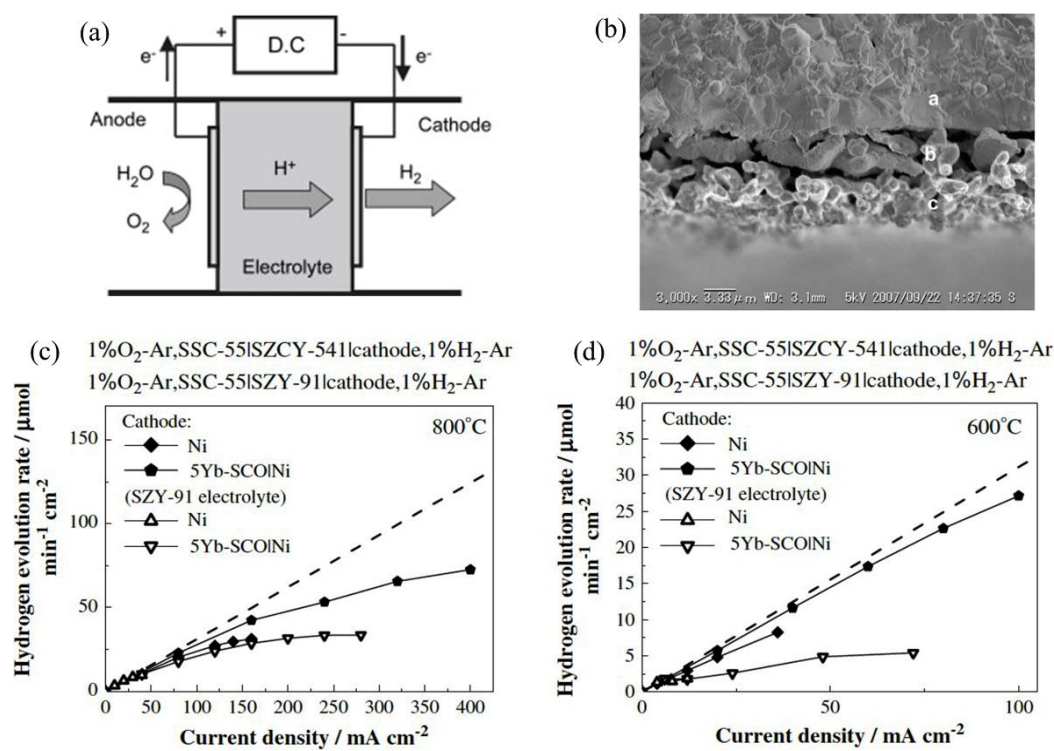

**Figure 11.** (**a**) Operating principle of steam electrolysis cells based on proton-conducting electrolytes (**b**) SEM images of the cross section of SCYb interlayer SZCY-541 electrolyte (top), SCYb interlayer (middle) and nickel electrode (bottom) (**c**,**d**) hydrogen evolution rate of the steam electrolysis cell with SZCY-541 electrolyte, SSC-55 anode, nickel cathode and SCYb interlayer at 800 °C (**c**) and 600 °C (**d**). Reproduced from Ref. [112]. Copyright © 2009 Elsevier.

The performance of a hydrogen pump consisting of proton-conducting oxide $SrZr_{0.8}In_{0.2}O_{3-\alpha}$ under various environments and temperature were investigated by Tanaka et al. (2008) [96]. Performance evaluation of the hydrogen pump was done by passing direct current within the temperature range 673–873 K. Wet argon and hydrogen gas were used as a fed gas at a pressure 1 kPa to the anode side. Results showed that mixed conduction (electron-hole, protonic, oxygen ion) occurred through $SrZr_{0.8}In_{0.2}O_{3-\alpha}$ electrolyte at a higher temperature, but with the increase in temperature, voltage and proton transport numbers decreased under water vapor electrolysis. The maximum proton transport number was observed at 823 K and started to decrease with temperature. Conversely, with the presence of hydrogen on the anode side, the proton transport number of oxygen ions increased with increasing temperature. It was concluded that a higher temperature was not necessary to enhance hydrogen pump performance, and a low temperature is practically favorable to reduce the migration of oxygen ions, leading to lower oxide degradation. The authors also studied the impact of oxygen partial pressure on the cathode side on hydrogen pump performance. The pressure was maintained over the range of 0.02 Pa to 1 kPa. It was found that the performance of the hydrogen pumps was enhanced with increasing oxygen partial pressure, which can be attributed to the enhancement of proton transport numbers (above $10^2$ Pa) and reduction in voltage. Furthermore, the comparison of the performance of hydrogen pumps based on $SrZr_{0.8}In_{0.2}O_{3-\alpha}$ and $CaZr_{0.9}In_{0.1}O_{3-\alpha}$ was studied under water vapor electrolysis at 873 K. Even though both showed mixed conduction at elevated temperatures, proton transport numbers and electrical resistance were lower for the $SrZr_{0.8}In_{0.2}O_{3-\alpha}$- than the $CaZr_{0.9}In_{0.1}O_{3-\alpha}$-based hydrogen pump. In addition, Figure 1 clearly shows that for 1.5 V, the protonic current of $SrZr_{0.8}In_{0.2}O_{3-\alpha}$ was 37 mA, whereas for $CaZr_{0.9}In_{0.1}O_{3-\alpha}$, it was 0.41 mA. In addition, the hydrogen evolution rate was higher for $CaZr_{0.9}In_{0.1}O_{3-\alpha}$. These results indicate that proton conductivity is the major parameter to enhance the performance of the hydrogen pump.

Sakai et al. (2009) [92] used palladium as an electrode to enhance the performance of the hydrogen pump. Since proton-conducting electrolyte SZY-91 was used, hydrogen was directly produced in the cathode without any separation process. The electrolyte was prepared by solid-state reaction, and palladium electrodes were prepared by a sputtering method to measure the hydrogen evolution rate and screen-printing method to measure the morphological effect of the electrode. In this paper, the electrodes were designed without a three-phase boundary (TPB) to improve the performance, which is a different technique compared to the authors' previous study [113]. This was possible due to the palladium's special property, i.e., hydrogen permeability. Since hydrogen ions can easily penetrate the porous palladium electrodes, the authors were able to design the electrodes without TPB, which improved the pump's performance. The experiment showed that overpotentials at cathode and anode using sputtered palladium electrodes were significantly reduced compared to platinum electrodes. However, the overpotential of the cathode did not reduce as much as that of the anode, so future research is needed to improve the performance of the palladium cathode. When pasted palladium was used as an electrode, cathode overpotential showed a constant nature upon the sintering temperature change. On the contrary, the overpotential of anode sintered at 900 °C showed a higher value while anode sintered at 1400 °C showed almost zero overpotential. From the electrode morphology, the authors explained that in the case of 1400 °C sintering temperature, the electrode was dense and uniformly covered with the electrolyte, which increased the contact area. This indicates that the rate-determining step (RDS) of the anode reaction takes place at the interface of the electrolyte/electrode. Partial pressure of hydrogen did not affect the anode performance. On the other hand, anode sintered at 900 °C was porous, so overpotential was high.

Fukada et al. (2009) [114] studied fuel cells with proton-conducting $SrZr_{0.9}Yb_{0.1}O_{3-\alpha}$ electrolyte to produce hydrogen directly from methane reliably and effectively. They investigated the RDS of mass hydrogen transfer from the anode through the electrolyte to the cathode for the proper design of the hydrogen pump. The experiment was tested with a ceramic tube made of $SrZr_{0.9}Yb_{0.1}O_{3-\alpha}$ and porous $Ni/SiO_2$ electrodes because they are not prone to react with $CO_2$ or CO, and tubes with Ni electrodes are more cost-effective than the platinum electrode. As the moist methane ($CH_4 + H_2O$) is supplied to the porous $Ni/SiO_2$ anode, the $CH_4$ dissociation reaction ($CH_4 + H_2O = CO + 2H_2$) occurs with the steam and produced $H_2$ gathers on the $Ni/SiO_2$ electrode surface. From the Arrhenius plot of electrolyte and electrode resistance, it was noticed that polarization resistance was larger than the ohmic resistance, and the resistance related to the $CH_4$ dissociation reaction was the RDS of the hydrogen transfer process from anode to cathode. Besides, reaction resistance at the cathode compartment was found to be negligible in the transfer process, which led the author to conclude that the rate-determining step was in the anode compartment. On the contrary, at the temperature range over 700 °C, the rate-determining step was found to be the proton transfer through the $SrZr_{0.9}Yb_{0.1}O_{3-\alpha}$ electrolyte, and the activation energy for proton transfer was 31.5 kJ/mol. It was also found that when the temperature was below 700 °C, the resistance related to the reaction at the cathode regulated the transfer rate of hydrogen.

Chen et al. (2012) [115] investigated the mechanism of proton conduction across the grain boundary of 5 mol% yttrium-doped polycrystalline strontium zirconate (p-SZY5), used as a hydrogen pump. This electrolyte is chemically stable compared to barium zirconate [116] and showed the highest proton conductivity at ~5 mol% concentration [117–120]. Before the experiment, the authors assumed a potential barrier at the grain boundary, which restricts proton transport, and the grain boundary acts as a Schottky junction. From the Arrhenius plot of average grain boundary and bulk conductivities, it can be noted that the activation energy for average grain boundary conductivity (1.22 eV) was almost twice compared to the activation energy for bulk conductivity (0.59 eV). Besides, average grain boundary conductivity was significantly lower than the bulk conductivity (almost 4–5 orders of magnitude) over the temperature range. Therefore, it verifies the authors' assumption of limiting the protonic current across the grain boundary in p-SZY5. Furthermore, from the I-V characteristics of the grain boundary, the authors noticed that grain-boundary current increases nonlinearly with the increase of grain boundary voltage, which is almost similar to the properties of back-to-back Schottky junction. This proves that due to the presence of the Schottky barrier, the internal proton transport in p-SZY5 is reduced. Table 4 summarizes the zirconate-based hydrogen pumps and their relevant parameters.

Tanaka et al. (2010) [97] proposed a hydrogen pump consisting of $SrZr_{0.9}Yb_{0.1}O_{3-\alpha}$ to extract hydrogen from a gas mixture. To simulate the gas mixture present in a hydrogen processing unit, argon gas carrying hydrogen and methane at 0.1 kPa and water vapor at 1.2 kPa was passed into the system at 973 K, 85 cm³/min. The measurements of gases (H₂, CH₄) at the inlet and outlet of the anode were made by a thermal conductivity detector (TCD) and hydrogen flame ionization detector (FID), respectively. The hydrogen and water vapor concentration at the anode outlet was observed as 0.5% and 1%, respectively, while methane was not detected. This phenomenon was successfully demonstrated by the decomposition of methane and part of water vapor into hydrogen gas. Considering the hydrogen pump composed of 0.5 kPa $H_2$ and 1 kPa water vapor, it was found that with the increasing current density (up to 1.1 mA/cm²) hydrogen evolution rate increased linearly along with the theoretical value. After 1.1 mA/cm², it increased nonlinearly but became much smaller than the theoretical values, which depicted the migration of other charge carriers to the proton conductor. However, until 0.4 V, 1.1 mA/cm², hydrogen extraction to the cathode was executed by hydrogen pumping while the proton number (PN) was unity. Then, hydrogen extraction was executed by water vapor electrolysis, and the presence of oxygen was observed at 0.95 V (Figure 12a)In addition, Figure 12b shows gas concentration change at the outlet as a function of temperature [97]. Later, oxygen supplied electron-hole and oxide ions. These carriers are responsible for the reduced efficiency of the hydrogen pump and the reduction of electrolytes. This issue could be solved by reducing the operating temperature to less than 773 K. But proton conduction was also getting reduced at a low operating temperature which is a big challenge in using $SrZrO_3$-based proton conductors. Furthermore, during the investigation of the methane decomposition mechanism, it was observed that carbon did not deposit into the substrate. In addition, the following catalytic reaction (Equations (1) and (2)) [97] was postulated based on the experimental results (only for plated electrode), which shifted to the right side during hydrogen extraction and confirmed the usefulness of a hydrogen pump consisting of proton conductor in hydrogen recovery.

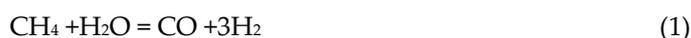

$$CH_4 + H_2O = CO + 3H_2 \tag{1}$$

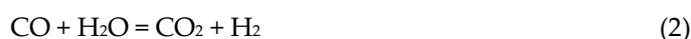

$$CO + H_2O = CO_2 + H_2 \tag{2}$$

Table 3. Zirconate based hydrogen sensors.

| Electrolyte Materials | Synthesis Method | Density | Electrode | Proton Conductivity (mS·cm⁻¹)/Temperature (°C) | Atmospheres | Hydrogen Concentration (unit)/TEMPERATURE (°C) | Working Temperature (°C) | Ref. |
|---|---|---|---|---|---|---|---|---|
| $CaZr_{0.9}In_{0.1}O_{3-\delta}$ | Solid state reaction | N/A | ZnO/Pt | $2.01 \times 10^{-4}$/700 | Wet Ar | 50–500 | 500–700 | [83] |
| | | | | $2.23 \times 10^{-4}$/700 | Air | 50–500 | 500–700 | |
| | | | | $2.41 \times 10^{-4}$/700 | 4000 ppm $H_2$/Ar | 50–500 | 500–700 | |
| $CaZr_{0.95}Mn_{0.05}O_{3-\alpha}$ | N/A | Relative density 98% | Pt/Pt | N/A | N/A | N/A | N/A | [88] |
| $CaZr_{0.9}In_{0.1}O_{3-\delta}$ | Solid state reaction | N/A | N/A | N/A | N/A | N/A | N/A | [89] |
| $CaZr_{0.95}Sc_{0.05}O_{3-\alpha}$ | Glycine-nitrate combustion | Relative density 99% | Yttria stabilized zirconia/Pt | N/A | N/A | N/A | 675–750 | [103] |
| $CaZr_{0.9}In_{0.1}O_{3-\delta}$ | Glycine-nitrate combustion | N/A | Pt | N/A | N/A | N/A | N/A | [121] |
| $CaZr_{0.9}In_{0.1}O_{3-\delta}$ | N/A | N/A | Pd | N/A | N/A | N/A | N/A | [110] |

Table 4. Zirconate-based hydrogen pump devices.

| Electrolyte Materials | Fabrication Method | Thickness (μm) | Electrode Type | Feed Gas at Anode | Feed Gas at Cathode | Sample Gas | Flow Rate (L/min) | Temperature (°C) | Ref. |
|---|---|---|---|---|---|---|---|---|---|
| $SrZr_{0.9}Yb_{0.1}O_{3-\alpha}$ | Solid state reaction | 500 | Pt | 1% $O_2$ + 99% Ar | 1% $H_2$ + 99% Ar | N/A | N/A | 600–800 | [112] |
| $SrZr_{0.9}Yb_{0.1}O_{3-\alpha}$ | Solid state reaction | 500 | Pd | N/A | N/A | 1% $H_2$ + 99% Ar | N/A | 800 | [92] |
| $SrZr_{0.9}Yb_{0.1}O_{3-\alpha}$ | N/A | N/A | Ni/SiO$_2$ | $CH_4$-Ar or $H_2$-Ar | N/A | $CH_4$ + $H_2O$ | N/A | 600–800 | [114] |
| $SrZr_{0.95}Yb_{0.05}O_{3-\alpha}$ | Pecchini method | 200–300 | N/A | N/A | N/A | N/A | N/A | N/A | [115] |
| $SrZr_{0.8}Yb_{0.2}O_{3-\alpha}$ | N/A | 720 | Pt | Wet $H_2$ gas or wet Ar | Wet Ar | N/A | 0.1 | N/A | [96] |
| $SrZr_{0.9}Yb_{0.1}O_{3-\alpha}$ | N/A | 1500 | Pt | $CH_4$ + $H_2$ | N/A | N/A | 0.1 | N/A | [97] |

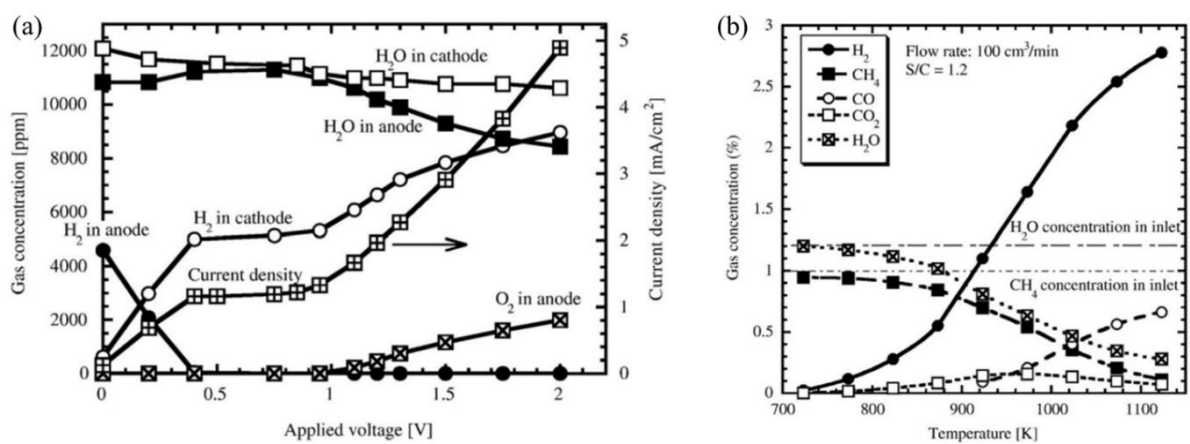

**Figure 12.** (**a**) Variation of current density and gas concentrations at the outlet of the anode and cathode as a function of applied voltage, (**b**) gas concentration change at the outlet as a function of temperature; here hydrogen, methane, carbon monoxide, carbon-di oxide, and water vapor are shown. Reproduced from Ref. [97]. Copyright © 2010 Elsevier.

## 5. Challenges and Prospects

Zirconated-based proton-conducting materials used as electrolytes in electrochemical devices face significant challenges. These challenges primarily involve performance, material selection, doping concentration, temperature, proton conductivity, cost, and water vapor concentration. In the following section, this paper presents the challenges and prospects of using zirconated materials in electrochemical devices.

### 5.1. Tritium Monitoring Systems

The common issue in $CaZr_{0.9}In_{0.1}O_{3-\alpha}$-based hydrogen pump performance is the fluctuation of hydrogen evolution rate at a high temperature, which eventually causes the use of multiple tubes to meet the commercial grade proportional counter [70]. However, several other factors that lower the hydrogen pump performance are the fluctuation in the concentration of $H_2O$-vapor in the anode compartment, an increase in the conductivity of electron-hole and oxide-ion, electrode conductivity, etc. In addition, high water vapor concentration in fed gas negatively affects the hydrogen pump performance, such as hydrogen recovery rate decreases with increasing $H_2O$-vapor concentration which was an issue [71]. Although the $H_2$-evolution rate using proton-conducting oxides increases with high temperature and voltage, low-temperature proton-conducting oxides are more desirable. In this regard, $SrZr_{0.9}Yb_{0.1}O_{3-\alpha}$ was found most suitable to extract maximum hydrogen at low temperatures, which is advantageous. Nevertheless, many ceramic tubes were required to obtain desired hydrogen evolution rate in the tritium monitoring system. Another issue in the application of proton-conducting ceramics with a pasted platinum electrode was the instantaneous evolution of water vapor on the cathode side after applying voltage (Figure 13) [97].

In water vapor electrolysis with a closed-loop system, hydrogen concentration was five times the water vapor concentration at the anode side after 50 min. It should be 10 times after 10 min in an actual application, which is a big challenge for a tritium monitoring system [122]. The introduction of a proportional type tritium monitor can solve this issue by detecting low-concentration tritium. Moreover, the concentration of tritium is increased with the increasing current density while the concentration is in the range of two orders in magnitude [69]. The use of wet argon gas as fed gas was proved as a useful method to decontaminate the tritium from the membrane separator [43].

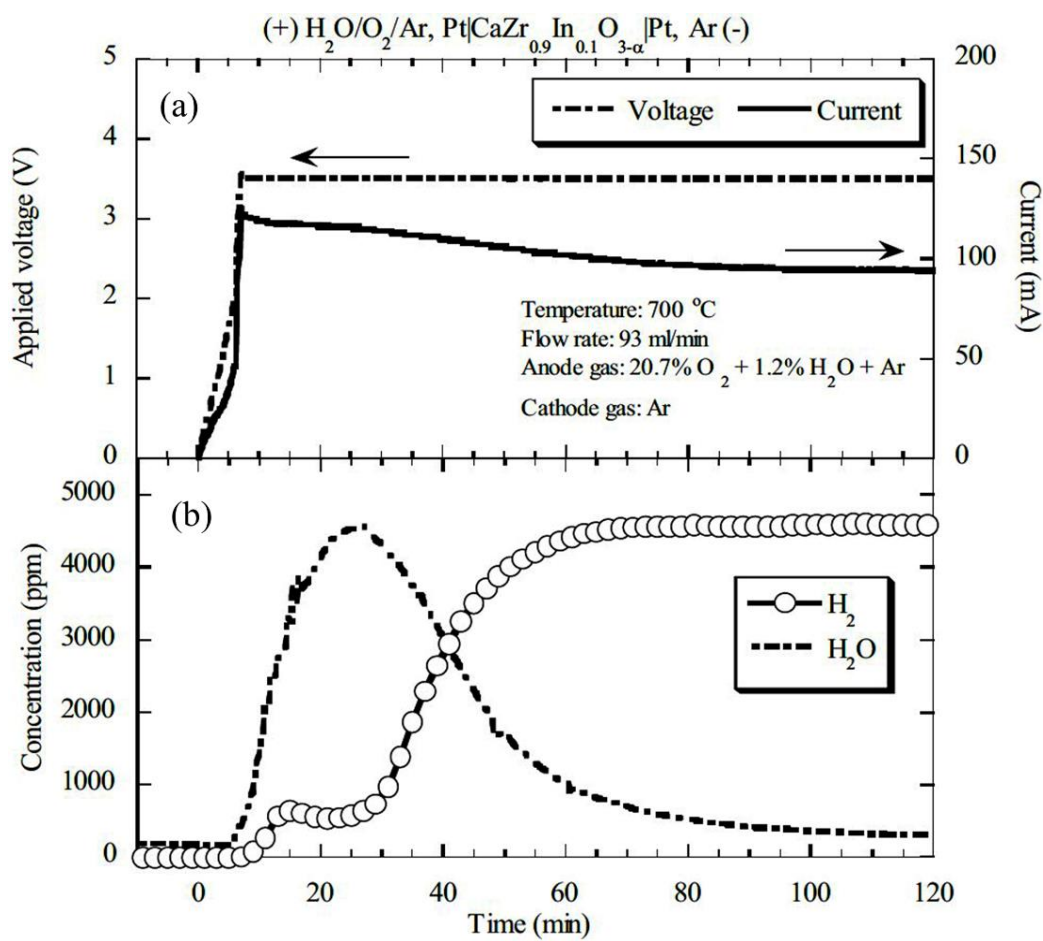

**Figure 13.** Time evolution of hydrogen and water vapor at the cathode outlet under the wet atmosphere containing oxygen; (**a**) applied voltage and current, (**b**) concentration of water vapor and hydrogen. Reproduced from Ref. [97]. Copyright © 2005 Taylor & Francis.

### 5.2. Tritium Recovery System

The current density is a factor that represents the performance of a tube-type proton conducting oxide. Enhancement of current density depends on a better preparation procedure for the electrode. In addition, the hydrogen evolution rate increases with current density under vacuum on the cathode side at a high temperature (973 K) after applying direct current. It was found that using an electroless platinum electrode instead of pasted electrode provides enhanced hydrogen evolution rate and current in the hydrogen pump. However, only the current increased in water vapor electrolysis, but the hydrogen evolution rate does not change, so to improve the hydrogen evolution rate, an electrode of better oxygen diffusivity is required. Posted electrode plays a vital role in this case due to its large pore size, but the vacuum condition on the cathode side must be confirmed [82]. High vacuum conditions may further enhance the hydrogen recovery rate, which needs to be investigated. A large electrode surface area is also required to increase the current density in applying tube-type proton conducting oxide. Otherwise, the requirement for the number of tubes would increase, which may increase the overall cost. Another parameter that should be considered to enhance the performance of proton-conducting oxide is proton conductivity instead of total conductivity. Increasing the oxygen partial pressure on the cathode side increases the proton conductivity and electrical resistance by preventing the reduction of proton-conducting oxide. It is also reported that the reduction of proton conducting oxide, e.g., $SrZr_{0.9}Yb_{0.1}O_{3-\alpha}$, can be prevented by reducing the operating temperature, but proton conduction is also reduced at a low operating temperature, which is a big challenge of using $SrZrO_3$-based proton conductors. Hence, $BaZrO_3$-based proton-conducting oxide can be a future candidate for operating at intermediate temperatures with higher proton conductivity [97].

### 5.3. Sensor Devices

Although using a zirconate-based proton conductor for hydrogen sensors has significant advantages, some challenges remain. The sensing properties depend mainly on the proton conductivity, which depends on the doping effects. However, when the calcination temperature is higher than 1400 °C, there is no improvement in doping found in the electrolyte. Even at 1550 °C, the doping effect degrades significantly [89]. Limiting current hydrogen sensors are the current-type hydrogen sensor that consists of a proton conductor with high chemical stability and proton conductivity [123,124]. Recent studies have reported that the limiting current is increased with increasing voltages across the electrode in the hydrogen sensor, and also partial pressure is proportional to the limiting current [123–125].

Zirconate electrochemical hydrogen sensor systems often use potentiometric, amperometric, and impedimetric mechanisms for sensing. Amperometric methods are the most widely available in hydrogen sensors. However, amperometric sensors, for example, $CaZr_{0.95}Sc_{0.05}O_{3-\delta}$ electrolyte-based hydrogen sensors, are limited by electrotype–electrode interfaces [83]. This is because the sensor's performance depends not only on the zirconate electrolyte but also on the electrode used in the system. Therefore, further research needs to be done on the electrode material and its compatibility with the zirconate electrolytes. Even though ZnO [83], Pt [89], etc., have been used as electrode materials, their characterization remains a mystery, and therefore the maximum sensing performance remains unrealized in the scientific community. Moreover, the effects of these microstructures on the triple-phase boundary (gas–electrode–electrolyte) are not fully known [83]. One more limitation of amperometric sensors is that this sensor needs to be a reference electrode that acts as the control electrode. This is required because the hydrogen signal needs to be subtracted from this reference to get the actual signal (noise reduction).

However, finding a suitable zirconate material and concentration for the electrolyte is still challenging, and prospective researchers have a lot of potential to work in that area. Doping is one more critical area where scientists still need to optimize the doping concentration to get maximum hydrogen sensor performance. Indium, ytterbium, manganese, and scandium have been used as doping materials [83,88,103]; however, the field remains open for prospective new doping materials (and concentration) that can improve the sensor performance beyond the current level. Finally, our current hydrogen sensors' operating temperature range is unknown. The temperature likely affects the performance, and the performance at high temperatures has not been evaluated. Since doping the zirconates remains challenging at a higher temperature, so does hydrogen sensor performance [89]. The study of sensing characteristics at high temperatures remains open for prospective researchers.

### 5.4. Hydrogen Pumps

Hydrogen pumping technologies are essential in a nuclear fusion plant. These hydrogen pumps utilize the methods of electrochemical cell technologies where some researchers used zirconate ceramic materials as electrolytes [97] and platinum/palladium/nickel materials as electrodes [79,92,112]. The main drawback of these methods still lies with material selection. The electrolyte surface becomes narrow because of the three-phase boundary, resulting in low electrode (electrochemical cell) performance [92]. A study has reported that palladium is used as an electrode (instead of platinum) with strontium zirconate materials to minimize the effect of three-phase boundaries [92]. Still, finding the right electrode–electrolyte combination for maximum performance of hydrogen pumps is a crucial challenge for prospective researchers in this field. Some studies reported a few models to explain the boundary conditions. One model assumes that a potential barrier is present at the three-phase boundaries, which limits the protonic conductivity. Another model assumes that the grain boundary is like a Schottky junction and that the protonic current follows thermionic emissions [115]. More innovative models are required from prospective scientists to explain this characteristic so we can finally overcome this barrier.

Another critical challenge is the high overpotential of platinum-based electrodes when used with zirconate conductors [92]. High electrode overpotentials led to the poor efficiency of the electrolytes in the pump. The method of doping with electrolytes proved to be a feasible solution to resolve this problem. Sakai et al. reported that palladium instead of platinum can be used as electrodes in $SrZr_{0.9}Yb_{0.1}O_{3-\alpha}$ electrolytes to improve the performance of the hydrogen pump [92,126]. However, this cannot significantly reduce the problem of overpotential at the cathode.

The hydrogen pumps consisting of $SrZr_{0.9}Yb_{0.1}O_{3-\alpha}$ faced significant challenges in maintaining the theoretical hydrogen evolution rate at higher current flow due to the migration of other charge carriers to the proton conductor [97]. Furthermore, zirconate-based ceramics also lead to high electrolyte analysis and low hydrogen pumping characteristics. To overcome this challenge, prospective researchers should collectively focus on developing methods to innovate novel materials for electrode/electrolyte for electrochemical hydrogen pumps. Finally, more research is required to precisely pinpoint the rate-limiting step of the hydrogen pump.

## 6. Conclusions

The slow depletion of fuel sources and awareness of green alternatives drives the scientific community to focus their efforts on cleaner sources of energy and fuel alternatives. Electrochemical method-based energy sources have garnered considerable attention among researchers for their potential in clean and renewable energy sources. Compared to the other proton-conducting materials used in electrochemical methods as an electrolyte, zirconate-based perovskite proton-conducting materials are the most promising materials due to their excellent proton conductivity and chemical stability in a wide temperature range. In this paper, we have reviewed the existing literature on the potential of zirconate materials as an electrolyte in tritium sensors, tritium recovery systems, hydrogen sensor devices, and hydrogen pump devices. We have drawn the following conclusions based on the existing literature.

I.    $CaZr_{0.9}In_{0.1}O_{3-\alpha}$ and $SrZr_{0.9}Yb_{0.1}O_{3-\alpha}$ proton conductors are extensively used for tritium monitoring. These materials have minimal interference with other radioactive molecules and perform well in higher temperature ranges.

II.   $CaZr_{0.9}In_{0.1}O_{3-\alpha}$, $SrZr_{0.8}In_{0.2}O_{3-\alpha}$, and $SrZr_{0.9}Yb_{0.1}O_{3-\alpha}$ can be utilized in a tritium recovery system. Furthermore, since they have proton conductivity more than cerates, their performance as a tritium recovery system is superior.

III.  Using different concentrations of calcium zirconates and materials ($CaZr_{0.9}In_{0.1}O_{3-\delta}$, $CaZr_{0.95}Sc_{0.05}O_{3-\delta}$, $CaZr_{1-x}Mn_xO_{3-d}$), researchers were able to successfully implement an effective hydrogen sensor device that is capable of operating in an extensive temperature range and without the need for a reference electrode.

IV.   $CaZr_{0.9}In_{0.1}O_{3-\alpha}$, $SrZr_{0.9}Y_{0.1}O_{3-\alpha}$, and $SrZr_{0.9}Yb_{0.1}O_{3-\alpha}$ electrolyte zirconate materials have been used as electrochemical hydrogen pumps. These materials have excellent hydrogen extraction properties and can minimize the rate-limiting step.

Using zirconate materials in electrochemical cells presents unique challenges to prospective researchers. Researchers over the years have come up with innovative computational and laboratory approaches to address these challenges. The challenges in the fields mentioned above are described as follows:

I.    Zirconate-based tritium monitoring system faces the challenge of hydrogen rate fluctuation. In addition, the real-time concentration is sometimes unreliable because of low temporal resolution. Decontamination of the membrane separator is an additional challenge for the zirconate electrolyte tritium sensor.

II.   In a tritium recovery system, the main challenge is the operating temperature of the zirconate materials. Although alternate materials are adequate to address the issue, finding the perfect materials remains challenging.

III.  Challenges in hydrogen sensor devices include electrode/electrolyte material selection, doping concentration, and temperature dependence in performance.

IV.   In electrochemical hydrogen pumps, material selection still poses a considerable challenge. Problems associated with overpotentials are a massive problem in anodic electrode materials. Moreover, there is a requirement for an innovative model with three-phase boundary conditions that will explain the correct emission and conductivity.

In this paper, we have laid out the utility of zirconate-based proton conductors in terms of tritium monitor, tritium recovery, a hydrogen sensor device, and hydrogen pumps. We have also pointed out the challenges and prospects of zirconate materials, which give a path to perspective researchers to make a meaningful contribution in this field.

**Author Contributions:** Conceptualization, M.K.H. and K.H.; methodology, M.K.H.; software, M.K.H.; validation, M.K.H., R.C.D. and K.H.; formal analysis, M.K.H., H.B., M.F.R., M.H.K.R., and R.C.D.; investigation, M.K.H.; resources, M.K.H. and K.H.; data curation, M.K.H., S.M.K.H. and T.E.; writing—original draft preparation, M.K.H., S.M.K.H., M.I.H. and T.E.; writing—review and editing, M.K.H., M.I.H., R.C.D., and K.H.; visualization, M.K.H., H.B., M.F.R., M.H.K.R., and K.H.; supervision, M.K.H. and K.H.; project administration, M.K.H. and K.H.; All authors have read and agreed to the published version of the manuscript.

**Funding:** This research did not receive any specific grant from funding agencies in the public, commercial, or not-for-profit sectors.

**Data Availability Statement:** Not applicable.

**Acknowledgments:** We acknowledge the MS and PhD students who initially supported us to organize the content of the manuscript.

**Conflicts of Interest:** The authors declare no conflicts of interest.

## References

1. Sharma, P.; Said, Z.; Kumar, A.; Nižetić, S.; Pandey, A.; Hoang, A.T.; Huang, Z.; Afzal, A.; Li, C.; Le, A.T.; et al. Recent Advances in Machine Learning Research for Nanofluid-Based Heat Transfer in Renewable Energy System. *Energy Fuels* **2022**, *36*, 6626–6658. https://doi.org/10.1021/acs.energyfuels.2c01006.
2. Siram, O.; Sahoo, N.; Saha, U.K. Changing Landscape of India's Renewable Energy and the Contribution of Wind Energy. *Clean. Eng. Technol.* **2022**, *8*, 100506. https://doi.org/10.1016/j.clet.2022.100506.
3. Sircar, A.; Tripathi, G.; Bist, N.; Shakil, K.A.; Sathiyanarayanan, M. *Emerging Technologies for Sustainable and Smart Energy*; CRC Press: Boca Raton, FL, USA, 2022; ISBN 9781003307402.
4. Chia, S.R.; Nomanbhay, S.; Ong, M.Y.; Shamsuddin, A.H.B.; Chew, K.W.; Show, P.L. Renewable Diesel as Fossil Fuel Substitution in Malaysia: A Review. *Fuel* **2022**, *314*, 123137. https://doi.org/10.1016/j.fuel.2022.123137.
5. Karakoc, T.H.; Colpan, C.O.; Ekici, S.; Yetik, O. Promising Fuels and Green Energy Technologies for Aviation. *Int. J. Green Energy* **2022**. https://doi.org/10.1080/15435075.2022.2082253.
6. Gogoi, P.; Tudu, B.; Saikia, P. Hydrogen Fuel: Clean Energy Production Technologies. In *Status and Future Challenges for Non-conventional Energy Sources Volume 1*; Springer: Berlin/Heidelberg, Germany, 2022; pp. 133–154.
7. Khan, M.; Das, R.C.; Casey, J.; Reese, B.L.; Akintunde, B.; Pathak, A.K. Near Room Temperature Magnetocaloric Properties in Ni Deficient $(Mn_{0.525}Fe_{0.5})Ni_{0.975}Si_{0.95}Al_{0.05}$. *AIP Adv.* **2022**, *12*, 035227. https://doi.org/10.1063/9.0000294.
8. Hossain, M.K.; Raihan, G.A.; Akbar, M.A.; Kabir Rubel, M.H.; Ahmed, M.H.; Khan, M.I.; Hossain, S.; Sen, S.K.; Jalal, M.I.E.; El-Denglawey, A. Current Applications and Future Potential of Rare Earth Oxides in Sustainable Nuclear, Radiation, and Energy Devices: A Review. *ACS Appl. Electron. Mater.* **2022**, *4*, 3327–3353. https://doi.org/10.1021/acsaelm.2c00069.
9. Andrews, J.; Rezaei Niya, S.M.; Ojha, R. Electrochemical Hydrogen Storage in Porous Carbons with Acidic Electrolytes: Uncovering the Potential. *Curr. Opin. Electrochem.* **2022**, *31*, 100850. https://doi.org/10.1016/j.coelec.2021.100850.
10. Sedighi, F.; Ghiyasiyan-Arani, M.; Behpour, M. Ternary Nanocomposites of $Ce_2W_2O_9/CoWO_4/Porous$ Carbon; Design, Structural Study and Electrochemical Hydrogen Storage Application. *Fuel* **2022**, *310*, 122218. https://doi.org/10.1016/j.fuel.2021.122218.
11. Hossain, M.K.; Rubel, M.H.K.; Akbar, M.A.; Ahmed, M.H.; Haque, N.; Rahman, M.F.; Hossain, J.; Hossain, K.M. A Review on Recent Applications and Future Prospects of Rare Earth Oxides in Corrosion and Thermal Barrier Coatings, Catalysts, Tribological, and Environmental Sectors. *Ceram. Int.* **2022**, *48*, 32588–32612. https://doi.org/10.1016/j.ceramint.2022.07.220.
12. Kreuer, K.D. Proton-Conducting Oxides. *Annu. Rev. Mater. Res.* **2003**, *33*, 333–359. https://doi.org/10.1146/annurev.matsci.33.022802.091825.
13. Colomban, P. Proton Conductors and Their Applications: A Tentative Historical Overview of the Early Researches. *Solid State Ionics* **2019**, *334*, 125–144. https://doi.org/10.1016/j.ssi.2019.01.032.
14. Duan, C.; Huang, J.; Sullivan, N.; O'Hayre, R. Proton-Conducting Oxides for Energy Conversion and Storage. *Appl. Phys. Rev.* **2020**, *7*, 011314. https://doi.org/10.1063/1.5135319.

15. Leonard, K.; Druce, J.; Thoreton, V.; Kilner, J.A.; Matsumoto, H. Exploring Mixed Proton/Electron Conducting Air Electrode Materials in Protonic Electrolysis Cell. *Solid State Ion.* **2018**, *319*, 218–222. https://doi.org/10.1016/j.ssi.2018.02.016.

16. Hossain, M.K.; Chanda, R.; El-Denglawey, A.; Emrose, T.; Rahman, M.T.; Biswas, M.C.; Hashizume, K. Recent Progress in Barium Zirconate Proton Conductors for Electrochemical Hydrogen Device Applications: A Review. *Ceram. Int.* **2021**, *47*, 23725–23748. https://doi.org/10.1016/j.ceramint.2021.05.167.

17. Hussain, S.; Li, Y. Review of Solid Oxide Fuel Cell Materials: Cathode, Anode, and Electrolyte. *Energy Transit.* **2020**, *4*, 113–126. https://doi.org/10.1007/s41825-020-00029-8.

18. Harada, K.; Tanii, R.; Matsushima, H.; Ueda, M.; Sato, K.; Haneda, T. Effects of Water Transport on Deuterium Isotope Separation during Polymer Electrolyte Membrane Water Electrolysis. *Int. J. Hydrogen Energy* **2020**, *45*, 31389–31395. https://doi.org/10.1016/j.ijhydene.2020.08.256.

19. Adhikari, S.; Fernando, S. Hydrogen Membrane Separation Techniques. *Ind. Eng. Chem. Res.* **2006**, *45*, 875–881. https://doi.org/10.1021/ie050644l.

20. Li, Y.; Kappis, K.; Papavasiliou, J.; Fu, Z.; Chen, L.; Li, H.; Vlachos, D.E.; Avgouropoulos, G. Insights on the Electrochemical Performance of a Molten Proton Conductor Fuel Cell with Internal Methanol Reformer. *J. Power Sources* **2022**, *542*, 231813. https://doi.org/10.1016/j.jpowsour.2022.231813.

21. Bonanos, N. Perovskite Proton Conductor. In *Encyclopedia of Applied Electrochemistry*; Springer: New York, NY, USA, 2014; pp. 1514–1520.

22. Pei, K.; Zhou, Y.; Xu, K.; Zhang, H.; Ding, Y.; Zhao, B.; Yuan, W.; Sasaki, K.; Choi, Y.; Chen, Y.; et al. Surface Restructuring of a Perovskite-Type Air Electrode for Reversible Protonic Ceramic Electrochemical Cells. *Nat. Commun.* **2022**, *13*, 2207. https://doi.org/10.1038/s41467-022-29866-5.

23. Rubel, M.H.K.; Mitro, S.K.; Hossain, M.K.; Hossain, K.M.; Rahaman, M.M.; Hossain, J.; Mondal, B.K.; Akter, A.; Rahman, M.F.; Ahmed, I.; et al. First-Principles Calculations to Investigate Physical Properties of Single-Cubic ($Ba_{0.82}K_{0.18}$)($Bi_{0.53}Pb_{0.47}$)$O_3$ Novel Perovskite Superconductor. *Mater. Today Commun.* **2022**, *33*, 104302. https://doi.org/10.1016/j.mtcomm.2022.104302.

24. Rubel, M.H.K.; Hossain, M.A.; Hossain, M.K.; Hossain, K.M.; Khatun, A.A.; Rahaman, M.M.; Ferdous Rahman, M.; Hossain, M.M.; Hossain, J. First-Principles Calculations to Investigate Structural, Elastic, Electronic, Thermodynamic, and Thermoelectric Properties of $CaPd_3B_4O_{12}$ (B = Ti, V) Perovskites. *Results Phys.* **2022**, *42*, 105977. https://doi.org/10.1016/j.rinp.2022.105977.

25. de Souza, E.C.C.; Muccillo, R. Properties and Applications of Perovskite Proton Conductors. *Mater. Res.* **2010**, *13*, 385–394. https://doi.org/10.1590/S1516-14392010000300018.

26. Zhou, M.; Liu, Z.; Chen, M.; Zhu, Z.; Cao, D.; Liu, J. Electrochemical Performance and Chemical Stability of Proton-conducting $BaZr_{0.8-x}Ce_xY_{0.2}O_{3-\delta}$ Electrolytes. *J. Am. Ceram. Soc.* **2022**, *105*, 5711–5724. https://doi.org/10.1111/jace.18500.

27. Zhou, D.; Zhou, T.; Tian, Y.; Zhu, X.; Tu, Y. Perovskite-Based Solar Cells: Materials, Methods, and Future Perspectives. *J. Nanomater.* **2018**, *2018*, 8148072. https://doi.org/10.1155/2018/8148072.

28. Yang, Y.; Ling, X.; Qiu, W.; Bian, J.; Zhang, X.; Chen, Q. Surface-Enhanced Raman Scattering Spectroscopy Reveals the Phonon Softening of Yttrium-Doped Barium Zirconate Thin Films. *J. Phys. Chem. C* **2022**, *126*, 10722–10728. https://doi.org/10.1021/acs.jpcc.2c01906.

29. Nur Syafkeena, M.A.; Zainor, M.L.; Hassan, O.H.; Baharuddin, N.A.; Othman, M.H.D.; Tseng, C.-J.; Osman, N. Review on the Preparation of Electrolyte Thin Films Based on Cerate-Zirconate Oxides for Electrochemical Analysis of Anode-Supported Proton Ceramic Fuel Cells. *J. Alloys Compd.* **2022**, *918*, 165434. https://doi.org/10.1016/j.jallcom.2022.165434.

30. Peltzer, D.; Múnera, J.; Cornaglia, L. Study of the Sorption Properties of Alkali Zirconate-Based Sorbents at High Temperature in the Presence of Water and Low $CO_2$ Concentration. *J. Alloys Compd.* **2022**, *895*, 162419. https://doi.org/10.1016/j.jallcom.2021.162419.

31. Rashid, N.L.R.M.; Samat, A.A.; Jais, A.A.; Somalu, M.R.; Muchtar, A.; Baharuddin, N.A.; Wan Isahak, W.N.R. Review on Zirconate-Cerate-Based Electrolytes for Proton-Conducting Solid Oxide Fuel Cell. *Ceram. Int.* **2019**, *45*, 6605–6615. https://doi.org/10.1016/j.ceramint.2019.01.045.

32. Chen, M.; Zhou, M.; Liu, Z.; Liu, J. A Comparative Investigation on Protonic Ceramic Fuel Cell Electrolytes $BaZr_{0.8}Y_{0.2}O_{3-\delta}$ and $BaZr_{0.1}Ce_{0.7}Y_{0.2}O_{3-\delta}$ with NiO as Sintering Aid. *Ceram. Int.* **2022**, *48*, 17208–17216. https://doi.org/10.1016/j.ceramint.2022.02.278.

33. Yamanaka, S.; Fujikane, M.; Hamaguchi, T.; Muta, H.; Oyama, T.; Matsuda, T.; Kobayashi, S.; Kurosaki, K. Thermophysical Properties of $BaZrO_3$ and $BaCeO_3$. *J. Alloys Compd.* **2003**, *359*, 109–113. https://doi.org/10.1016/S0925-8388(03)00214-7.

34. Kurosaki, K.; Adachi, J.; Maekawa, T.; Yamanaka, S. Thermal Conductivity Analysis of $BaUO_3$ and $BaZrO_3$ by Semiempirical Molecular Dynamics Simulation. *J. Alloys Compd.* **2006**, *407*, 49–52. https://doi.org/10.1016/j.jallcom.2005.06.045.

35. Borland, H.; Llivina, L.; Colominas, S.; Abellà, J. Proton Conducting Ceramics for Potentiometric Hydrogen Sensors for Molten Metals. *Fusion Eng. Des.* **2013**, *88*, 2431–2435. https://doi.org/10.1016/j.fusengdes.2013.05.055.

36. Tanaka, M.; Ohshima, T. Recovery of Hydrogen from Gas Mixture by an Intermediate-Temperature Type Proton Conductor. *Fusion Eng. Des.* **2010**, *85*, 1038–1043. https://doi.org/10.1016/j.fusengdes.2010.01.007.

37. Pergolesi, D.; Fabbri, E.; D'Epifanio, A.; Di Bartolomeo, E.; Tebano, A.; Sanna, S.; Licoccia, S.; Balestrino, G.; Traversa, E. High Proton Conduction in Grain-Boundary-Free Yttrium-Doped Barium Zirconate Films Grown by Pulsed Laser Deposition. *Nat. Mater.* **2010**, *9*, 846–852. https://doi.org/10.1038/nmat2837.

38. Assabumrungrat, S.; Sangtongkitcharoen, W.; Laosiripojana, N.; Arpornwichanop, A.; Charojrochkul, S.; Praserthdam, P. Effects of Electrolyte Type and Flow Pattern on Performance of Methanol-Fuelled Solid Oxide Fuel Cells. *J. Power Sources* **2005**, *148*, 18–23. https://doi.org/10.1016/j.jpowsour.2005.01.034.

39. Sun, W.; Liu, M.; Liu, W. Chemically Stable Yttrium and Tin Co-Doped Barium Zirconate Electrolyte for Next Generation High Performance Proton-Conducting Solid Oxide Fuel Cells. *Adv. Energy Mater.* **2013**, *3*, 1041–1050. https://doi.org/10.1002/aenm.201201062.

40. Hossain, M.K.; Biswas, M.C.; Chanda, R.K.; Rubel, M.H.K.; Khan, M.I.; Hashizume, K. A Review on Experimental and Theoretical Studies of Perovskite Barium Zirconate Proton Conductors. *Emergent Mater.* **2021**, *4*, 999–1027. https://doi.org/10.1007/s42247-021-00230-5.

41. Zhang, W.; Hu, Y.H. Progress in Proton-conducting Oxides as Electrolytes for Low-temperature Solid Oxide Fuel Cells: From Materials to Devices. *Energy Sci. Eng.* **2021**, *9*, 984–1011. https://doi.org/10.1002/ese3.886.

42. Schwandt, C.; Fray, D.J. The Titanium/Hydrogen System as the Solid-State Reference in High-Temperature Proton Conductor-Based Hydrogen Sensors. *J. Appl. Electrochem.* **2006**, *36*, 557–565. https://doi.org/10.1007/s10800-005-9108-5.

43. Tanaka, M.; Sugiyama, T.; Ohshima, T.; Yamamoto, I. Extraction of Hydrogen and Tritium Using High-Temperature Proton Conductor for Tritium Monitoring. *Fusion Sci. Technol.* **2011**, *60*, 1391–1394. https://doi.org/10.13182/FST11-A12690.

44. Miller, J.M.; Bokwa, S.R.; Macdonald, D.S.; Verrall, R.A. Tritium Recovery from Lithium Zirconate Spheres. *Fusion Technol.* **1991**, *19*, 996–999. https://doi.org/10.13182/FST91-A29472.

45. Hossain, M.K.; Tamura, H.; Hashizume, K. Visualization of Hydrogen Isotope Distribution in Yttrium and Cobalt Doped Barium Zirconates. *J. Nucl. Mater.* **2020**, *538*, 152207. https://doi.org/10.1016/j.jnucmat.2020.152207.

46. Hossain, M.K.; Hashizume, K.; Hatano, Y. Evaluation of the Hydrogen Solubility and Diffusivity in Proton-Conducting Oxides by Converting the PSL Values of a Tritium Imaging Plate. *Nucl. Mater. Energy* **2020**, *25*, 100875. https://doi.org/10.1016/j.nme.2020.100875.

47. Hossain, M.K.; Iwasa, T.; Hashizume, K. Hydrogen Isotope Dissolution and Release Behavior in Y-doped BaCeO₃. *J. Am. Ceram. Soc.* **2021**, *104*, 6508–6520. https://doi.org/10.1111/jace.18035.

48. Hossain, M.K.; Yamamoto, T.; Hashizume, K. Effect of Sintering Conditions on Structural and Morphological Properties of Y- and Co-Doped BaZrO₃ Proton Conductors. *Ceram. Int.* **2021**, *47*, 27177–27187. https://doi.org/10.1016/j.ceramint.2021.06.138.

49. Hossain, M.K.; Yamamoto, T.; Hashizume, K. Isotopic Effect of Proton Conductivity in Barium Zirconates for Various Hydrogen-Containing Atmospheres. *J. Alloys Compd.* **2022**, *903*, 163957. https://doi.org/10.1016/j.jallcom.2022.163957.

50. Han, D.; Liu, X.; Bjørheim, T.S.; Uda, T. Yttrium-Doped Barium Zirconate-Cerate Solid Solution as Proton Conducting Electrolyte: Why Higher Cerium Concentration Leads to Better Performance for Fuel Cells and Electrolysis Cells. *Adv. Energy Mater.* **2021**, *11*, 2003149. https://doi.org/10.1002/aenm.202003149.

51. Liu, Y.; Zhang, W.; Wang, B.; Sun, L.; Li, F.; Xue, Z.; Zhou, G.; Liu, B.; Nian, H. Theoretical and Experimental Investigations on High Temperature Mechanical and Thermal Properties of BaZrO₃. *Ceram. Int.* **2018**, *44*, 16475–16482. https://doi.org/10.1016/j.ceramint.2018.06.064.

52. Draber, F.M.; Ader, C.; Arnold, J.P.; Eisele, S.; Grieshammer, S.; Yamaguchi, S.; Martin, M. Publisher Correction: Nanoscale Percolation in Doped BaZrO₃ for High Proton Mobility. *Nat. Mater.* **2020**, *19*, 577–577. https://doi.org/10.1038/s41563-020-0654-3.

53. Hossain, M.K.; Hashizume, K. Dissolution and Release Behavior of Hydrogen Isotopes from Barium-Zirconates. *Proc. Int. Exch. Innov. Conf. Eng. Sci.* **2020**, *6*, 34–39. https://doi.org/10.5109/4102460.

54. Perrichon, A.; Jedvik Granhed, E.; Romanelli, G.; Piovano, A.; Lindman, A.; Hyldgaard, P.; Wahnström, G.; Karlsson, M. Unraveling the Ground-State Structure of BaZrO₃ by Neutron Scattering Experiments and First-Principles Calculations. *Chem. Mater.* **2020**, *32*, 2824–2835. https://doi.org/10.1021/acs.chemmater.9b04437.

55. Abdalla, A.M.; Hossain, S.; Radenahmad, N.; Petra, P.M.I.; Somalu, M.R.; Rahman, S.M.H.; Eriksson, S.G.; Azad, A.K. Synthesis and Characterization of Sm₁₋ₓZrₓFe₁₋ᵧMgᵧO₃ (x, y = 0.5, 0.7, 0.9) as Possible Electrolytes for SOFCs. *Key Eng. Mater.* **2018**, *765*, 49–53. https://doi.org/10.4028/www.scientific.net/KEM.765.49.

56. Dai, H.; Kou, H.; Wang, H.; Bi, L. Electrochemical Performance of Protonic Ceramic Fuel Cells with Stable BaZrO₃-Based Electrolyte: A Mini-Review. *Electrochem. Commun.* **2018**, *96*, 11–15. https://doi.org/10.1016/j.elecom.2018.09.001.

57. Park, K.-Y.; Seo, Y.; Kim, K.B.; Song, S.-J.; Park, B.; Park, J.-Y. Enhanced Proton Conductivity of Yttrium-Doped Barium Zirconate with Sinterability in Protonic Ceramic Fuel Cells. *J. Alloys Compd.* **2015**, *639*, 435–444. https://doi.org/10.1016/j.jallcom.2015.03.168.

58. Yoo, Y.; Lim, N. Performance and Stability of Proton Conducting Solid Oxide Fuel Cells Based on Yttrium-Doped Barium Cerate-Zirconate Thin-Film Electrolyte. *J. Power Sources* **2013**, *229*, 48–57. https://doi.org/10.1016/j.jpowsour.2012.11.094.

59. Shafi, S.P.; Bi, L.; Boulfrad, S.; Traversa, E. Yttrium and Nickel Co-Doped BaZrO₃ as a Proton-Conducting Electrolyte for Intermediate Temperature Solid Oxide Fuel Cells. *ECS Trans.* **2015**, *68*, 503–508. https://doi.org/10.1149/06801.0503ecst.

60. Bi, L.; Traversa, E. Synthesis Strategies for Improving the Performance of Doped-BaZrO₃ Materials in Solid Oxide Fuel Cell Applications. *J. Mater. Res.* **2014**, *29*, 1–15. https://doi.org/10.1557/jmr.2013.205.

61. Dai, H. Proton Conducting Solid Oxide Fuel Cells with Chemically Stable BaZr₀.₇₅Y₀.₂Pr₀.₀₅O₃₋δ Electrolyte. *Ceram. Int.* **2017**, *43*, 7362–7365. https://doi.org/10.1016/j.ceramint.2017.02.090.

62. Liu, Y.; Ran, R.; Tade, M.O.; Shao, Z. Structure, Sinterability, Chemical Stability and Conductivity of Proton-Conducting BaZr₀.₆M₀.₂Y₀.₂O₃₋δ Electrolyte Membranes: The Effect of the M Dopant. *J. Memb. Sci.* **2014**, *467*, 100–108. https://doi.org/10.1016/j.memsci.2014.05.020.

63. Loureiro, F.J.A.; Nasani, N.; Reddy, G.S.; Munirathnam, N.R.; Fagg, D.P. A Review on Sintering Technology of Proton Conducting BaCeO₃-BaZrO₃ Perovskite Oxide Materials for Protonic Ceramic Fuel Cells. *J. Power Sources* **2019**, *438*, 226991. https://doi.org/10.1016/j.jpowsour.2019.226991.

64. Rajendran, S.; Thangavel, N.K.; Ding, H.; Ding, Y.; Ding, D.; Reddy Arava, L.M. Tri-Doped BaCeO₃–BaZrO₃ as a Chemically Stable Electrolyte with High Proton-Conductivity for Intermediate Temperature Solid Oxide Electrolysis Cells (SOECs). *ACS Appl. Mater. Interfaces* **2020**, *12*, 38275–38284. https://doi.org/10.1021/acsami.0c12532.

65. Jurewicz, K.; Frackowiak, E.; Béguin, F. Towards the Mechanism of Electrochemical Hydrogen Storage in Nanostructured Carbon Materials. *Appl. Phys. A* **2004**, *78*, 981–987. https://doi.org/10.1007/s00339-003-2418-8.

66. Korotcenkov, G.; Han, S.D.; Stetter, J.R. Review of Electrochemical Hydrogen Sensors. *Chem. Rev.* **2009**, *109*, 1402–1433. https://doi.org/10.1021/CR800339K/ASSET/IMAGES/MEDIUM/CR-2008-00339K_0039.GIF.

67. Bouwman, P. Electrochemical Hydrogen Compression (EHC) Solutions for Hydrogen Infrastructure. *Fuel Cells Bull.* **2014**, *2014*, 12–16. https://doi.org/10.1016/S1464-2859(14)70149-X.

68. Keçebaş, A.; Kayfeci, M.; Bayat, M. Electrochemical Hydrogen Generation. In *Solar Hydrogen Production*; Elsevier: Amsterdam, The Netherlands, 2019; pp. 299–317.

69. Tanaka, M.; Sugiyama, T. Development of a Tritium Monitor Combined with an Electrochemical Tritium Pump Using a Proton Conducting Oxide. *Fusion Sci. Technol.* **2015**, *67*, 600–603. https://doi.org/10.13182/FST14-T89.

70. Tanaka, M.; Katahira, K.; Asakura, Y.; Uda, T.; Iwahara, H.; Yamamoto, I. Hydrogen Extraction Characteristics of Proton-Conducting Ceramics under a Wet Air Atmosphere for a Tritium Stack Monitor. *J. Nucl. Sci. Technol.* **2004**, *41*, 1013–1017. https://doi.org/10.1080/18811248.2004.9726325.

71. Tanaka, M.; Asakura, Y.; Uda, T.; Katahira, K.; Tsuji, N.; Iwahara, H. Hydrogen Enrichment by Means of Electrochemical Hydrogen Pump Using Proton-Conducting Ceramics for a Tritium Stack Monitor. *Fusion Eng. Des.* **2006**, *81*, 1371–1377. https://doi.org/10.1016/j.fusengdes.2005.08.075.

72. Tanaka, M.; Asakura, Y.; Uda, T.; Katahira, K.; Iwahara, H.; Tsuji, N.; Yamamoto, I. Studies on Hydrogen Extraction Characteristics of Proton-Conducting Ceramics and Their Applications to a Tritium Recovery System and a Tritium Monitor. *Fusion Sci. Technol.* **2005**, *48*, 51–54. https://doi.org/10.13182/FST05-A878.

73. Hossain, M.K. Study on Hydrogen Isotopes Behavior in Proton Conducting Zirconates and Rare Earth Oxides. Ph.D.Thesis, Kyushu University, Fukuoka, Japan, 2021.

74. Hossain, M.K.; Ahmed, M.H.; Khan, M.I.; Miah, M.S.; Hossain, S. Recent Progress of Rare Earth Oxides for Sensor, Detector, and Electronic Device Applications: A Review. *ACS Appl. Electron. Mater.* **2021**, *3*, 4255–4283. https://doi.org/10.1021/acsaelm.1c00703.

75. Hossain, M.K.; Hashizume, K.; Jo, S.; Kawaguchi, K.; Hatano, Y. Hydrogen Isotope Dissolution and Release Behavior of Rare Earth Oxides. *Fusion Sci. Technol.* **2020**, *76*, 553–566. https://doi.org/10.1080/15361055.2020.1728173.

76. Hossain, M.K.; Kawaguchi, K.; Hashizume, K. Isotopic Effect of Proton Conductivity in Gadolinium Sesquioxide. *Fusion Eng. Des.* **2021**, *171*, 112555. https://doi.org/10.1016/j.fusengdes.2021.112555.

77.  Khalid Hossain, M.; Kawaguchi, K.; Hashizume, K. Conductivity of Gadolinium (III) Oxide ($Gd_2O_3$) in Hydrogen-Containing Atmospheres. *Proc. Int. Exch. Innov. Conf. Eng. Sci.* **2020**, *6*, 1–6. https://doi.org/10.5109/4102455.

78.  Hossain, M.K.; Kawaguchi, K.; Hashizume, K. Protonic Conductivity and Isotope Dependency in Rare-Earth Gadolinium Oxide. In Proceedings of the 22nd Cross Straits Symposium on Energy and Environmental Science and Technology (CSS-EEST22), Fukuoka, Japan, 2-3 December 2020; Kyushu University: Fukuoka, Japan, 2020; pp. 15–16.

79.  Fukada, S.; Suemori, S.; Onoda, K. Overall Conductivity and Electromotive Force of $SrZr_{0.9}Yb_{0.1}O_{3-a}$ Cell System Supplied with Moist CH 4. *J. Nucl. Sci. Technol.* **2007**, *44*, 1324–1329. https://doi.org/10.1080/18811248.2007.9711378.

80.  Xia, T.; He, C.; Yang, H.; Zhao, W.; Yang, L. Hydrogen Extraction Characteristics of High-Temperature Proton Conductor Ceramics for Hydrogen Isotopes Purification and Recovery. *Fusion Eng. Des.* **2014**, *89*, 1500–1504. https://doi.org/10.1016/j.fusengdes.2014.04.078.

81.  Tanaka, M.; Katahira, K.; Asakura, Y.; Uda, T.; Iwahara, H.; Yamamoto, I. Effect of Plated Platinum Electrode on Hydrogen Extraction Performance Using $CaZrO_3$-Based Proton-Conducting Ceramic for Tritium Recovery System. *J. Nucl. Sci. Technol.* **2004**, *41*, 95–97. https://doi.org/10.1080/18811248.2004.9715464.

82.  Tanaka, M. Extraction of Hydrogen into Vacuum by Electrochemical Hydrogen Pump for Hydrogen Isotope Recovery. *Fusion Eng. Des.* **2012**, *87*, 1065–1069. https://doi.org/10.1016/j.fusengdes.2012.02.082.

83.  Dai, L.; Wang, L.; Shao, G.; Li, Y. A Novel Amperometric Hydrogen Sensor Based on Nano-Structured ZnO Sensing Electrode and $CaZr_{0.9}In_{0.1}O_{3-\delta}$ Electrolyte. *Sens. Actuators B Chem.* **2012**, *173*, 85–92. https://doi.org/10.1016/j.snb.2012.06.012.

84.  Ma, J.; Zhou, Y.; Bai, X.; Chen, K.; Guan, B.-O. High-Sensitivity and Fast-Response Fiber-Tip Fabry–Pérot Hydrogen Sensor with Suspended Palladium-Decorated Graphene. *Nanoscale* **2019**, *11*, 15821–15827. https://doi.org/10.1039/C9NR04274A.

85.  Ding, Y.; Li, Y.; Huang, W. Influence of Grain Interior and Grain Boundaries on Transport Properties of Scandium-doped Calcium Zirconate. *J. Am. Ceram. Soc.* **2020**, *103*, 2653–2662. https://doi.org/10.1111/jace.16968.

86.  Zeba, I.; Ramzan, M.; Ahmad, R.; Shakil, M.; Rizwan, M.; Rafique, M.; Sarfraz, M.; Ajmal, M.; Gillani, S.S.A. First-Principles Computation of Magnesium Doped $CaZrO_3$ Perovskite: A Study of Phase Transformation, Bandgap Engineering and Optical Response for Optoelectronic Applications. *Solid State Commun.* **2020**, *313*, 113907. https://doi.org/10.1016/j.ssc.2020.113907.

87.  Dunyushkina, L.A.; Khaliullina, A.S.; Meshcherskikh, A.N.; Pankratov, A.A. Sintering and Conductivity of Sc-Doped $CaZrO_3$ with $Fe_2O_3$ as a Sintering Aid. *Ceram. Int.* **2021**, *47*, 10565–10573. https://doi.org/10.1016/j.ceramint.2020.12.168.

88.  Okuyama, Y.; Nagamine, S.; Nakajima, A.; Sakai, G.; Matsunaga, N.; Takahashi, F.; Kimata, K.; Oshima, T.; Tsuneyoshi, K. Proton-Conducting Oxide with Redox Protonation and Its Application to a Hydrogen Sensor with a Self-Standard Electrode. *RSC Adv.* **2016**, *6*, 34019–34026. https://doi.org/10.1039/C5RA23560J.

89.  Chen, J.; Wu, S.; Zhang, F.; Lü, S.; Mao, Y. Calcination Temperature Dependence of Synthesis Process and Hydrogen Sensing Properties of In-Doped $CaZrO_3$. *Mater. Chem. Phys.* **2016**, *172*, 87–97. https://doi.org/10.1016/j.matchemphys.2015.12.064.

90.  Tong, Y.; Wang, Y.; Cui, C.; Wang, S.; Xie, B.; Peng, R.; Chen, C.; Zhan, Z. Preparation and Characterization of Symmetrical Protonic Ceramic Fuel Cells as Electrochemical Hydrogen Pumps. *J. Power Sources* **2020**, *457*, 228036. https://doi.org/10.1016/j.jpowsour.2020.228036.

91.  Matsumoto, H. Extraction and Production of Hydrogen Using High-Temperature Proton Conductor. *Solid State Ionics* **2002**, *152–153*, 715–720. https://doi.org/10.1016/S0167-2738(02)00414-9.

92.  Sakai, T.; Matsumoto, H.; Yamamoto, R.; Kudo, T.; Okada, S.; Watanabe, M.; Hashimoto, S.; Takamura, H.; Ishihara, T. Performance of Palladium Electrode for Electrochemical Hydrogen Pump Using Strontium-Zirconate-Based Proton Conductors. *Ionics* **2009**, *15*, 665–670. https://doi.org/10.1007/s11581-009-0365-x.

93.  Tanaka, M.; Katahira, K.; Asakura, Y.; Uda, T.; Iwahara, H.; Yamamoto, I. Hydrogen Extraction Using One-End Closed Tube Made of $CaZrO_3$-Based Proton-Conducting Ceramic for Tritium Recovery System. *J. Nucl. Sci. Technol.* **2004**, *41*, 61–67. https://doi.org/10.1080/18811248.2004.9715458.

94.  Tanaka, M.; Asakura, Y.; Uda, T. Performance of the Electrochemical Hydrogen Pump of a Proton-Conducting Oxide for the Tritium Monitor. *Fusion Eng. Des.* **2008**, *83*, 1414–1418. https://doi.org/10.1016/j.fusengdes.2008.06.038.

95.  Uchida, H. Relation between Proton and Hole Conduction in $SrCeO_3$-Based Solid Electrolytes under Water-Containing Atmospheres at High Temperatures. *Solid State Ion.* **1983**, *11*, 117–124. https://doi.org/10.1016/0167-2738(83)90048-6.

96. Tanaka, M.; Asakura, Y.; Uda, T. Experimental Study on Electrochemical Hydrogen Pump of SrZrO$_3$-Based Oxide. *Fusion Sci. Technol.* **2008**, *54*, 479–482. https://doi.org/10.13182/FST08-A1858.

97. Tanaka, M.; Katahira, K.; Asakura, Y.; Ohshima, T. Hydrogen Pump Using a High-Temperature Proton Conductor for Nuclear Fusion Engineering Applications. *Solid State Ion.* **2010**, *181*, 215–218. https://doi.org/10.1016/j.ssi.2009.01.020.

98. Han, J.; Wen, Z.; Zhang, J.; Wu, X.; Lin, B. Electrical Conductivity of Fully Densified Nano CaZr$_{0.90}$In$_{0.10}$O$_{3-\delta}$ Ceramics Prepared by a Water-Based Gel Precipitation Method. *Solid State Ion.* **2009**, *180*, 154–159. https://doi.org/10.1016/j.ssi.2008.12.016.

99. Li, Y.; Ding, Y.; Cui, S.; Wang, C. Preparation and Electrical Properties of Sc-Doped CaZrO$_3$. *Acta Metall. Sin.* **2013**, *48*, 575–578. https://doi.org/10.3724/SP.J.1037.2011.00776.

100. Yang, W.; Li, G.; Sui, Z. Coprecipitating Synthesis and Impedance Study of CaZr$_{1-x}$In$_x$O$_{3-\sigma}$. *J. Mater. Sci. Lett.* **1998**, *17*, 241–243. https://doi.org/10.1023/A:1006548731332.

101. Matsumoto, H. Hydrogen Isotope Cell and Its Application to Hydrogen Isotope Sensing. *Solid State Ion.* **2000**, *136–137*, 173–177. https://doi.org/10.1016/S0167-2738(00)00308-8.

102. Matsumoto, H.; Takeuchi, K.; Iwahara, H. Electromotive Force of Hydrogen Isotope Cell with a High Temperature Proton-conducting Solid Electrolyte CaZr$_{0.90}$In$_{0.10}$O$_{3-\alpha}$. *J. Electrochem. Soc.* **2019**, *146*, 1486–1491. https://doi.org/10.1149/1.1391791.

103. Kalyakin, A.S.; Lyagaeva, J.G.; Chuikin, A.Y.; Volkov, A.N.; Medvedev, D.A. A High-Temperature Electrochemical Sensor Based on CaZr$_{0.95}$Sc$_{0.05}$O$_{3-\delta}$ for Humidity Analysis in Oxidation Atmospheres. *J. Solid State Electrochem.* **2019**, *23*, 73–79. https://doi.org/10.1007/s10008-018-4108-7.

104. Zhou, M.; Ahmad, A. Sol–Gel Processing of In-Doped CaZrO$_3$ Solid Electrolyte and the Impedimetric Sensing Characteristics of Humidity and Hydrogen. *Sens. Actuators B Chem.* **2008**, *129*, 285–291. https://doi.org/10.1016/j.snb.2007.08.022.

105. Iwahara, H.; Asakura, Y.; Katahira, K.; Tanaka, M. Prospect of Hydrogen Technology Using Proton-Conducting Ceramics. *Solid State Ion.* **2004**, *168*, 299–310. https://doi.org/10.1016/j.ssi.2003.03.001.

106. Kang, B.S.; Heo, Y.W.; Tien, L.C.; Norton, D.P.; Ren, F.; Gila, B.P.; Pearton, S.J. Hydrogen and Ozone Gas Sensing Using Multiple ZnO Nanorods. *Appl. Phys. A* **2005**, *80*, 1029–1032. https://doi.org/10.1007/s00339-004-3098-8.

107. Wang, J.X.; Sun, X.W.; Yang, Y.; Huang, H.; Lee, Y.C.; Tan, O.K.; Vayssieres, L. Hydrothermally Grown Oriented ZnO Nanorod Arrays for Gas Sensing Applications. *Nanotechnology* **2006**, *17*, 4995–4998. https://doi.org/10.1088/0957-4484/17/19/037.

108. Liao, L.; Lu, H.B.; Li, J.C.; He, H.; Wang, D.F.; Fu, D.J.; Liu, C.; Zhang, W.F. Size Dependence of Gas Sensitivity of ZnO Nanorods. *J. Phys. Chem. C* **2007**, *111*, 1900–1903. https://doi.org/10.1021/jp065963k.

109. Shao, C.; Chang, Y.; Long, Y. High Performance of Nanostructured ZnO Film Gas Sensor at Room Temperature. Sensors Actuators B Chem. 2014, 204, 666–672, doi:10.1016/j.snb.2014.08.003.

110. Ohshima, T.; Kondo, M.; Tanaka, M.; Muroga, T.; Sagara, A. Hydrogen Transport in Molten Salt Flinak Measured by Solid Electrolyte Sensors with Pd Electrode. *Fusion Eng. Des.* **2010**, *85*, 1841–1846. https://doi.org/10.1016/j.fusengdes.2010.06.008.

111. Kurita, N.; Fukatsu, N.; Ito, K.; Ohashi, T. Protonic Conduction Domain of Indium-doped Calcium Zirconate. *J. Electrochem. Soc.* **2019**, *142*, 1552–1559. https://doi.org/10.1149/1.2048611.

112. Sakai, T.; Matsushita, S.; Matsumoto, H.; Okada, S.; Hashimoto, S.; Ishihara, T. Intermediate Temperature Steam Electrolysis Using Strontium Zirconate-Based Protonic Conductors. *Int. J. Hydrogen Energy* **2009**, *34*, 56–63. https://doi.org/10.1016/j.ijhydene.2008.10.011.

113. Sakai, T.; Matsumoto, H.; Kudo, T.; Yamamoto, R.; Niwa, E.; Okada, S.; Hashimoto, S.; Sasaki, K.; Ishihara, T. High Performance of Electroless-Plated Platinum Electrode for Electrochemical Hydrogen Pumps Using Strontium-Zirconate-Based Proton Conductors. *Electrochim. Acta* **2008**, *53*, 8172–8177. https://doi.org/10.1016/j.electacta.2008.06.013.

114. Fukada, S.; Suemori, S.; Onoda, K. Rate Determining Step of Direct Hydrogen and Electricity Production in SrZr$_{0.9}$Yb$_{0.1}$O$_{3-a}$ Fuel Cell Supplied with CH$_4$+H$_2$O. *Energy Convers. Manag.* **2009**, *50*, 1249–1255. https://doi.org/10.1016/j.enconman.2009.01.023.

115. Chen, C.-T.; Kim, S.K.; Ibbotson, M.; Yeung, A.; Kim, S. Thermionic Emission of Protons across a Grain Boundary in 5 Mol% Y-Doped SrZrO$_3$, a Hydrogen Pump. *Int. J. Hydrogen Energy* **2012**, *37*, 12432–12437. https://doi.org/10.1016/j.ijhydene.2012.05.102.

116. Oyama, Y.; Kojima, A.; Li, X.; Cervera, R.B.; Tanaka, K.; Yamaguchi, S. Phase Relation in the BaO–ZrO$_2$–YO$_{1.5}$ System: Presence of Separate BaZrO$_3$ Phases and Complexity in Phase Formation. *Solid State Ion.* **2011**, *197*, 1–12. https://doi.org/10.1016/j.ssi.2011.06.006.

117. Müller, J.; Kreuer, K..; Maier, J.; Matsuo, S.; Ishigame, M. A Conductivity and Thermal Gravimetric Analysis of a Y-Doped SrZrO$_3$ Single Crystal. *Solid State Ion.* **1997**, *97*, 421–427. https://doi.org/10.1016/S0167-2738(97)00087-8.

118. Yajima, T.; Suzuki, H.; Yogo, T.; Iwahara, H. Protonic Conduction in SrZrO$_3$-Based Oxides. *Solid State Ion.* **1992**, *51*, 101–107. https://doi.org/10.1016/0167-2738(92)90351-O.

119. Huang, H.; Ishigame, M.; Shin, S. Protonic Conduction in the Single Crystals of Y-Doped SrZrO$_3$. *Solid State Ion.* **1991**, *47*, 251–255. https://doi.org/10.1016/0167-2738(91)90246-8.

120. Higuchi, T.; Tsukamoto, T.; Sata, N.; Hiramoto, K.; Ishigame, M.; Shin, S. Protonic Conduction in the Single Crystals of SrZr$_{0.95}$M$_{0.09}$O$_3$ (M=Y, Sc, Yb, Er). *Jpn. J. Appl. Phys.* **2001**, *40*, 4162–4163. https://doi.org/10.1143/JJAP.40.4162.

121. Kalyakin, A.S.; Lyagaeva, J.Y.; Volkov, A.N.; Medvedev, D.A. Unusual Oxygen Detection by Means of a Solid State Sensor Based on a CaZr$_{0.9}$In$_{0.1}$O$_{3-\delta}$ Proton-Conducting Electrolyte. *J. Electroanal. Chem.* **2019**, *844*, 23–26. https://doi.org/10.1016/j.jelechem.2019.05.003.

122. Asakura, Y.; Sugiyama, T.; Kawano, T.; Uda, T.; Tanaka, M.; Tsuji, N.; Katahira, K.; Iwahara, H. Application of Proton-Conducting Ceramics and Polymer Permeable Membranes for Gaseous Tritium Recovery. *J. Nucl. Sci. Technol.* **2004**, *41*, 863–870. https://doi.org/10.1080/18811248.2004.9715558.

123. Taniguchi, N.; Kuroha, T.; Nishimura, C.; Iijima, K. Characteristics of Novel BaZr$_{0.4}$Ce$_{0.4}$In$_{0.2}$O$_3$ Proton Conducting Ceramics and Their Application to Hydrogen Sensors. *Solid State Ion.* **2005**, *176*, 2979–2983. https://doi.org/10.1016/j.ssi.2005.09.035.

124. Yang, W.; Wang, L.; Li, Y.; Zhou, H.; He, Z.; Liu, H.; Dai, L. A Limiting Current Hydrogen Sensor Based on BaHf$_{0.8}$Fe$_{0.2}$O$_{3-\delta}$ Dense Diffusion Barrier and BaHf$_{0.7}$Sn$_{0.1}$In$_{0.2}$O$_{3-\delta}$ Protonic Conductor. *Ceram. Int.* **2022**, *48*, 22113–22123. https://doi.org/10.1016/j.ceramint.2022.04.199.

125. Wang, X.; Liu, T.; Yu, J.; Li, L.; Zhang, X. A New Application of CeZr$_1$-O$_2$ as Dense Diffusion Barrier in Limiting Current Oxygen Sensor. *Sens. Actuators B Chem.* **2019**, *285*, 391–397. https://doi.org/10.1016/j.snb.2019.01.086.

126. *Utilization of Hydrogen for Sustainable Energy and Fuels*; Van de Voorde, M., Ed.; De Gruyter: Berlin, Germany, 2021; ISBN 9783110596274.